\documentclass[12pt, nofootinbib]{article}

\usepackage{amssymb}
\usepackage{amsmath,bm}
\usepackage{amssymb}
\usepackage{graphicx}
\usepackage{amsfonts}         
\usepackage{fancybox}

\usepackage{enumitem}

\usepackage{slashed}

\usepackage[font=small,labelfont=bf]{caption}
\DeclareCaptionFont{tiny}{\tiny}
\usepackage{epstopdf}
\usepackage[usenames,dvipsnames]{xcolor}
\usepackage[pagebackref,  
bookmarks={false}, pdfauthor={John McGreevy}, pdftitle={Yay, physics!}]{hyperref}
\hypersetup{colorlinks=true, 
linkcolor=BrickRed, 
citecolor=Violet, 
filecolor=OliveGreen, 
urlcolor=RoyalBlue, 
filebordercolor={.8 .8 1}, 
urlbordercolor={.8 .8 0}}
\usepackage{soul}
\setstcolor{Red}

\definecolor{darkgreen}{rgb}{0,0.4,0}
\definecolor{darkred}{rgb}{0.4,0,0}
\definecolor{darkblue}{rgb}{0,0,0.4}
\definecolor{lightblue}{rgb}{.6,.6,0.9}

\def\redt{\textcolor{darkred}}

\definecolor{uglybrown}{rgb}{0.8,  0.7,  0.5}

\definecolor{palatinatepurple}{rgb}{0.41, 0.16, 0.38}
\definecolor{celebrationcolor}{rgb}{0.75,  0.0,  0.9}

\definecolor{shadecolor}{rgb}{0.90,0.90,0.90}
\definecolor{DVcolor}{rgb}{0.95,  0.5,  0.2}
\definecolor{lightbluemuons}{rgb}{0.0,.65,1.0}

\usepackage{wasysym}

\usepackage[yyyymmdd,hhmmss]{datetime}   
\usepackage{framed}  
 \usepackage{mdframed}
 \usepackage{wrapfig}
 \usepackage{yfonts}  

\def\parfig#1#2{
\parbox{#1\textwidth}
{\includegraphics[width=#1\textwidth]{#2}}
}

\numberwithin{equation}{section}

\renewcommand{\theequation}{\arabic{section}.\arabic{equation}}

\def\nd{{ \vphantom{\dagger}}}

\newcommand{\vev}[1]{\langle #1 \rangle}

\newlength{\extraspace}
\newlength{\extraspaces}
\def\be{\begin{equation}}
\def\ee{\end{equation}}

\newcommand{\bea}{\begin{eqnarray}}
\newcommand{\eea}{\end{eqnarray}}

\def\dbar{{\mathchar'26\mkern-12mu \dd}}

\def\eps{\epsilon}
\def\half{{1\over 2}}

\def\Im{{\rm Im\hskip0.1em}}

\def\bra#1{\left\langle#1\right|}
\def\ket#1{\left|#1\right\rangle}

\def\vev#1{\left\langle{#1}\right\rangle}

\def\CI{{\cal I}}

\def\CO{{\cal O}}

\def\sgn{{\rm sgn\ }}

\def\II{\relax{I\kern-.10em I}}

\def\IB{\relax{\rm I\kern-.18em B}}

\def\ID{\relax{\rm I\kern-.18em D}}
\def\IE{\relax{\rm I\kern-.18em E}}
\def\IF{\relax{\rm I\kern-.18em F}}
\def\IG{\relax\hbox{$\inbar\kern-.3em{\rm G}$}}
\def\IGa{\relax\hbox{${\rm I}\kern-.18em\Gamma$}}
\def\IH{\relax{\rm I\kern-.18em H}}
\def\II{\relax{\rm I\kern-.18em I}}
\def\IK{\relax{\rm I\kern-.18em K}}

\def\inbar{\,\vrule height1.5ex width.4pt depth0pt}

\def\IR{\mathbb{R}}

\def\gs{g_s}
\def\lp10{\ell_p^{10}}
\def\lp11{\ell_p^{11}}
\def\R11{R_{11}}

\def\frac#1#2{{#1 \over #2}}

\newdimen\tableauside\tableauside=1.0ex
\newdimen\tableaurule\tableaurule=0.4pt
\newdimen\tableaustep
\def\phantomhrule#1{\hbox{\vbox to0pt{\hrule height\tableaurule width#1\vss}}}
\def\phantomvrule#1{\vbox{\hbox to0pt{\vrule width\tableaurule height#1\hss}}}
\def\sqr{\vbox{%
  \phantomhrule\tableaustep
  \hbox{\phantomvrule\tableaustep\kern\tableaustep\phantomvrule\tableaustep}%
  \hbox{\vbox{\phantomhrule\tableauside}\kern-\tableaurule}}}
\def\squares#1{\hbox{\count0=#1\noindent\loop\sqr
  \advance\count0 by-1 \ifnum\count0>0\repeat}}
\def\tableau#1{\vcenter{\offinterlineskip
  \tableaustep=\tableauside\advance\tableaustep by-\tableaurule
  \kern\normallineskip\hbox
    {\kern\normallineskip\vbox
      {\gettableau#1 0 }%
     \kern\normallineskip\kern\tableaurule}%
  \kern\normallineskip\kern\tableaurule}}
\def\gettableau#1 {\ifnum#1=0\let\next=\null\else
  \squares{#1}\let\next=\gettableau\fi\next}

 \def\eqnn#1{\xdef #1{(\secsym\the\meqno)}\writedef{#1\leftbracket#1}%
 \global\advance\meqno by1\wrlabeL#1}
 \def\eqna#1{\xdef #1##1{\hbox{$(\secsym\the\meqno##1)$}}
 \writedef{#1\numbersign1\leftbracket#1{\numbersign1}}%
 \global\advance\meqno by1\wrlabeL{#1$\{\}$}}
 \def\eqn#1#2{\xdef #1{(\secsym\the\meqno)}\writedef{#1\leftbracket#1}%
 \global\advance\meqno by1$$#2\eqno#1\eqlabeL#1$$}

\global\newcount\itemno \global\itemno=0

\def\itemaut#1{\global\advance\itemno by1\noindent\item{\the\itemno.}#1}

\def\({\left(}
\def\){\right)}

\def\ii{{\bf i}}

\def\dd{\text{d}}

\def\HH{{\bf H}}

\def\lsim{\mathrel{\mathstrut\smash{\ooalign{\raise2.5pt\hbox{$<$}\cr\lower2.5pt\hbox{$\sim$}}}}}
\def\gsim{\mathrel{\mathstrut\smash{\ooalign{\raise2.5pt\hbox{$>$}\cr\lower2.5pt\hbox{$\sim$}}}}}

\def\overleftrightarrow#1{\vbox{\ialign{##\crcr
     $\leftrightarrow$\crcr\noalign{\kern-0pt\nointerlineskip}
     $\hfil\displaystyle{#1}\hfil$\crcr}}}
     
     \def\overleftarrow#1{\vbox{\ialign{##\crcr
     $\leftarrow$\crcr\noalign{\kern-0pt\nointerlineskip}
     $\hfil\displaystyle{#1}\hfil$\crcr}}}

\def\gU{\textsf{U}}

\def\gs{\text{gs}}

\newif{\ifeq}           
\eqtrue                 
\newcounter{lecturecounter}

\def\baselinestretch{1.1}
\renewcommand{\title}[1]{\vbox{\center\LARGE{#1}}\vspace{5mm}}
\renewcommand{\author}[1]{\vbox{\center#1}\vspace{5mm}}
\newcommand{\address}[1]{\vbox{\center\em#1}}

\renewcommand{\date}[1]{\vbox{\center#1}}

\usepackage{feynmp-auto}

\def\rchi{{\redt \chi}}
\def\sG{ \redt{\cal G}}

\begin{document}
\begin{titlepage}
\title{\huge Strange metal from local quantum chaos}

\author{Daniel Ben-Zion and John McGreevy}







\address{Department of Physics, University of California at San Diego, La Jolla, CA 92093, USA}

\begin{abstract}


How to make a model of a
non-Fermi-liquid metal
with efficient current dissipation
is a long-standing problem.
Results from holographic duality
suggest a framework where local critical fermionic degrees of freedom 
provide both a source of decoherence for the Landau quasiparticle,
and a sink for its momentum.
This leads us to study a Kondo lattice type model with SYK models in place of the spin impurities.
We find evidence for a stable phase at intermediate couplings.


\end{abstract}
\vfill

\today

\end{titlepage}

\setcounter{tocdepth}{1}    

\renewcommand{\baselinestretch}{0.5}\normalsize
\tableofcontents
\renewcommand{\baselinestretch}{1.1}
\normalsize

\section{Introduction}

How to make a model of a metal which is not a Fermi liquid, 
both in terms of the single-electron physics
and in terms of its transport properties,
is a long-standing problem in theoretical physics.
A general field-theoretic strategy 
to make a non-Fermi liquid metal (NFL)
is to couple a Fermi surface
to some other gapless degrees of freedom.  
If those modes are bosonic (such as gauge fields 
or fluctuations of an order parameter)\footnote
{For a review of the large literature, see \cite{SungSik2017}.}, the coupling 
must be (at least) trilinear, schematically $ \psi^\dagger \psi \phi$, 
and the Landau quasiparticle decays predominantly by emission of 
soft $\phi$ modes. 
This process does not change the current much;
in such models, therefore, the transport
lifetime is much longer than the single-particle lifetime.  
On the other hand, there seem to exist NFLs where the two timescales
are comparable, and have the same temperature dependence.
This suggests that there should be other ways to make a NFL.

Not long ago, some people \cite{Lee:2008xf, Liu:2009dm, Cubrovic:2009ye, Faulkner:2009wj, Faulkner:2010zz} were desperate enough to make progress
on this problem that they tried to use gauge/gravity duality: 
an exotic large-$N$ conformal field theory 
with a dual description in terms of Einstein gravity in one higher dimension
was subjected to a chemical potential
for a global $\gU(1)$ symmetry.\footnote{For a more leisurely discussion of these issues, see also \S5 of \cite{McGreevy:2016myw}.}  
The retarded Green's function of local fermionic operators in the resulting state
revealed a Fermi surface in momentum space, near which 
the self-energy behaved as a power-law in frequency: 
$$ G_\psi(\omega, k) \buildrel{\text{small $\omega$}}\over{\sim} 
 {1\over  \omega - v_F k_\perp - \redt{\mathcal G(\omega) }} $$
 with $\sG(\omega) \sim \omega^{2 \nu}$,
 indicative of a non-Fermi liquid metal.
The special case of $ \nu \to 1/2$, where $\sG \sim \omega \log \omega$, is the marginal Fermi liquid Green's function of \cite{VL8996}.

In \cite{Faulkner:2009wj}, the power-law behavior was traced
the region of the extra-dimensional geometry 
near the black-hole horizon.
With the benefit of some hindsight \cite{Faulkner:2009wj, Faulkner:2010tq}, the key 
feature of 
the near-horizon geometry of the black hole in this construction
is that it describes a $z=\infty$ fixed point: its fluctuations
are power-law in frequency, and essentially\footnote{In fact, as emphasized in \cite{Iqbal:2011in}, in the holographic construction described above, there is a weak, analytic dependence on the momentum.
The authors of \cite{Iqbal:2011in} call this `semi-local criticality'.  
This is a feature of the holographic strange metal construction that we will not reproduce.}
independent of momentum --
they are localized critical excitations.
Hence, when coupled to a Fermi surface, 
they are able to render incoherent the propagation
of the quasiparticles,
and at the same time absorb arbitrary amounts of their momentum.
Therefore, in a model where 
the quasiparticle decay is dominated
by scattering off these excitations, the transport lifetime
will equal the single-particle lifetime,
and the power law in the conductivity $\rho(T)$
will match that of the fermion self-energy, 
as in the marginal Fermi liquid phenomenology \cite{VL8996}.

The holographic construction summarized above, 
or even its `semi-holographic' reduction \cite{Faulkner:2009wj, Faulkner:2010tq},
have the drawback that the description 
of the $z=\infty$ fixed point is in terms
of a mysterious gravitational system, whose
dynamics is only under control in a limit $N\to \infty$
with infinitely many degrees of freedom at each point in space.    
Corrections to this limit require one to confront quantum gravity,
or at least the back-reaction of quantum effects on the geometry 
\cite{Allais:2012ye, Allais:2013lha}.  
It would be useful to replace the near-horizon
$AdS_2 \times \IR^2$
region of the geometry with 
a more tractable locally critical system.

Such local quantum criticality is a fascinating 
idea,
whose realization
is desirable also 
as a justification of dynamical mean field theory 
\cite{Georges:1996zz, Kotliar:2006zz}.
Such a fixed point is roughly a critical theory
at each point in space, 
and hence requires the participation of 
many degrees of freedom.
As explained in \cite{Jensen:2011su, Jensen:2016pah, Maldacena:2016upp, Almheiri:2016fws},
this intuition can be made precise 
by studying the dependence of the density of 
states on the energy.  Dimensional analysis requires
$$ {dn\over dE}(E) = e^{S_0} \delta(E) + e^{S_1} {1\over E} .$$
The first term represents a groundstate entropy $S_0$
and violates the Third Law of Thermodynamics.
The second term is not integrable
and requires the appearance of a new energy scale
which violates the $z=\infty$ scaling
and, as a consequence of this argument, cannot be disentangled from the low energy physics.
The holographic construction is most naturally interpreted
in the canonical ensemble, 
and at $T \ll \mu$ and leading order in $N$, 
gives an extensive entropy which remains nonzero at $T \to 0$,
suggesting a violation of the Third Law and 
the associated instabilities.
The low-energy fate of the construction is obscured 
since classical gravity requires $N \to \infty$ before $ T/\mu \to 0$,
and by the fact that the gravity construction
involves many degrees of freedom besides the Fermi surface.

This discussion motivates the study of 
more accessible constructions 
of $z=\infty$ fixed points, 
to which one might couple a Fermi surface.
With this in mind, we cannot 
avoid thinking about the SYK (Sachdev-Ye-Kitaev) model \cite{1993PhRvL..70.3339S, 2000PhRvL..85..840G, Kitaev-2015, Sachdev:2015efa},
which is a solvable model of local quantum criticality, 
and which has many features in common 
with (dilaton) gravity in $AdS_2$ \cite{2010JSMTE..11..022S, Sachdev:2010um, Kitaev-2015, Sachdev:2015efa}.
For our purposes of destroying quasiparticles, 
we require a $z=\infty$ fixed point with 
fermion operators carrying a conserved $\gU(1)$ charge.
Such a generalization of the SYK model
is provided in \cite{Sachdev:2015efa}:
\be\label{eq:HSYK} H_\text{SYK} = \sum_{ijkl}^N J_{ijkl} \rchi^\dagger_i \rchi^\dagger_j \rchi^\nd_k \rchi^\nd_l
~~~~~~~\overline{J_{ijkl}} = 0, ~\overline{J_{ijkl}^2} = 
\frac{ J^2}{2N^3}~~.\ee
Its low-energy physics should be similar to dilaton gravity 
plus electromagnetism in $AdS_2$.

A single SYK model has no notion of space, 
since each fermion talks to every other.
Since we are interested in the effects 
of the $z=\infty$ fixed point 
on the physics of the Fermi surface, 
we must introduce some notion of locality.  
Therefore, we consider a lattice
of SYK clusters, decoupled from each other at the outset.
Depicting a single `cluster' of 
complex fermions as \parfig{.1}{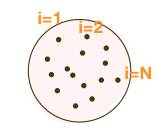},
a 1d implementation of the model can be 
illustrated 
as follows:
\begin{figure}[h] \begin{center}
\includegraphics[width=\textwidth]{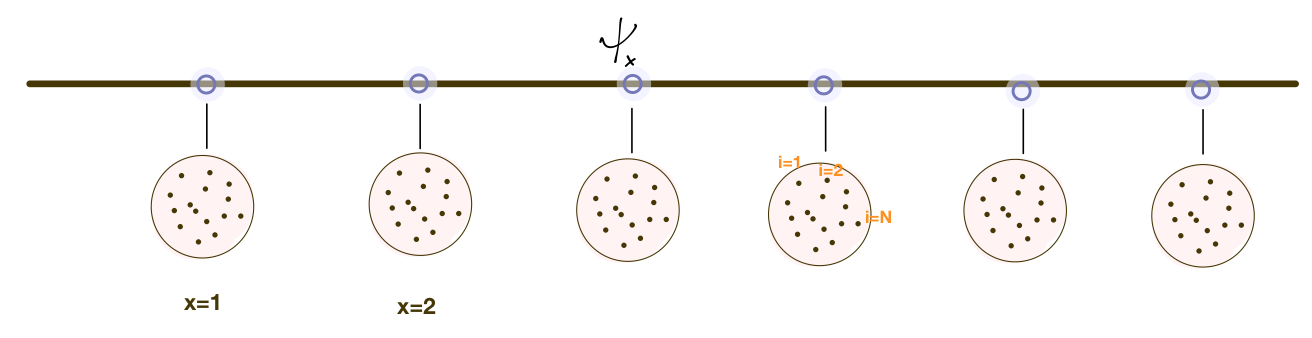}
\end{center}
\caption{A cartoon of the model we study in this paper.  
Each blob represents an independent SYK model.
The vertical edges represent the random couplings $g_{ix}$. 
The horizontal edges represent the translation-invariant 
hopping amplitudes $t$.
\label{fig:cluster-diagram}
}
\end{figure}

The model is a rather direct and crude 
discretization of the $AdS_2 \times \IR^d$ near-horizon
geometry of the extremal charged black hole in $AdS_{d+2}$.
Nevertheless, we will see that it reproduces
many of the features of interest of the holographic strange metal
of \cite{Faulkner:2009wj, Faulkner:2010zz}.

To be specific, the hamiltonian we will study is $H = H_0 + H_\text{int}$, with
$$ H_0 = -t\sum_{ \vev{xy} \in \text{lattice}}   \psi_x^\dagger \psi_y + h.c. 
+ \sum_{x \in \text{lattice}} H_{SYK}(\rchi_{xi}, J^x_{ijkl}) ,\quad H_\text{int} = \sum_{x,i} g^\nd_{ix} \psi^\dagger_x \rchi^\nd_{xi} + \textrm{h.c.} $$
where $\psi_x, \rchi_{xi}$ are complex canonical fermion annihilation operators,
with
$ \{ \psi^\dagger_x, \psi^\nd_y\} = \delta_{xy}$. Since $\psi$  form a Fermi surface under $H_0$, we refer to them as itinerant fermions. We occasionally refer to the $\rchi$ modes as cluster fermions. 
The couplings $g_{ix}$ are independently Gaussian:
$$\overline{ g_{ix}} = 0, ~\overline{ g_{ix} g_{jy }} = \delta_{ij} \delta_{xy} g^2/N.$$

There are some precedents for our study.  
The result of hybridizing conduction electrons with 
the SY (as opposed to SYK) model,
and its connection with holography, is studied in 
\cite{2010JSMTE..11..022S, Sachdev:2010um}. 
The model studied in this paper 
is simpler in that no fractionalization is required to 
write down the Hamiltonian.

The system we study here has some similarities
with models of heavy fermions, 
and in particular those devoted to 
understanding NFL behavior in those systems, 
such as, for example, \cite{miranda1996kondo}.
This paper solves a model 
of conduction electrons coupled to 
localized $f$-electrons by random 
hybridization terms.
The $f$-electrons have random site energies
and a uniform Hubbard $U$.  
The model is approximated using dynamical mean field theory.
There is a large literature studying
such heavy-fermion-like models using DMFT.
One goal of this work is to 
understand better the
local (momentum-independent) form of the self-energy 
assumed by the DMFT analysis.


Some related work
has also appeared during the overly long gestation of our project.
\cite{gu2017local} makes lattices of SYK clusters,
coupled by a less dangerous four-fermion coupling,
and studies the propagation of information.
\cite{banerjee2017solvable} studies the coupling of a single SYK cluster
to fermions which can hop (essentially in infinite dimensions) by 
the same kind of hybridization term we study; this model
lacks a notion of locality, however.
\cite{2017arXiv170305111H} couples non-locally several 
flavors of SYK clusters.
\cite{Chen:2017dav} studies the phase diagram of two clusters by quadratic terms.
Most recently and closest to our work, \cite{song2017strongly, Zhang:2017jvh, 2017arXiv171000842H} 
study a chain of SYK clusters
coupled by (random and non-random) quadratic links; although the starting point
does not have a Fermi surface, the resulting states of matter 
may be closely related to ours.
Studies of higher-dimensional generalizations of the SYK model, with various motivations, include
\cite{Berkooz:2016cvq, Turiaci:2017zwd, Berkooz:2017efq, Jian:2017jfl, Gu:2017ohj, Khveshchenko:2017mvj, Murugan:2017eto,Jian:2017tzg}.  In particular, \cite{Jian:2017tzg} realizes a bosonic analog of the semi-holographic construction
using SYK chains.

In the next section, we analyze the model at large $N$,
arriving at the same picture as in the semi-holographic models.
The advantage of having an explicit model of the $z \to \infty$ fixed point
is that we can 
analyze the extent to which the large-$N$ and low-energy limits commute.
In section \S\ref{sec:finite-N}, we analyze limits of the space of couplings 
and map out possible phase diagrams.
In section \S\ref{sec:qto2}, we attempt to
make the fixed point perturbative by continuing 
in the number $q$ of fermions participating in the SYK interactions.
In section \S\ref{sec:numerical-results}
we describe a DMRG study to decide between the possible 
phase diagrams proposed in \S\ref{sec:finite-N}.  

\begin{figure}[h] \begin{center}
\includegraphics[width=.4\textwidth]{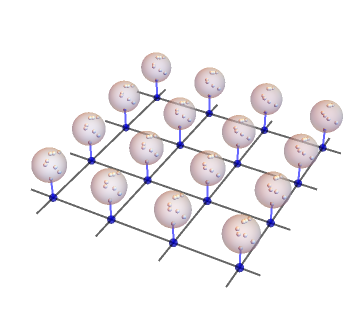}
\end{center}
\caption{
\label{fig-2d}
A diagram of the model in two spatial dimensions.  
The vertical (blue) bonds represent the random hybridization couplings $g_{xi}$.
The black horizontal bonds are the uniform hoppings, $t$. 
}
\end{figure}

\begin{figure}[h] \begin{center}
\includegraphics[width=.5\textwidth]{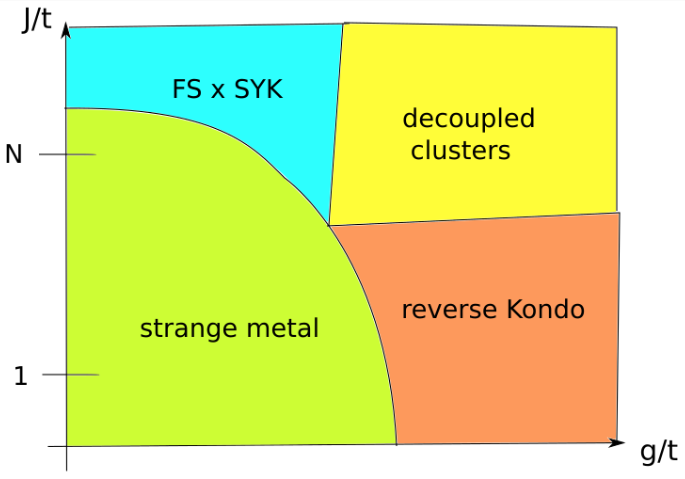}
\end{center}
\caption{
\label{fig:phase-diagram}
A (possibly optimistic) cartoon view of the proposed phase diagram.  
``reverse Kondo" refers to the regime
where one linear combination of the fermions in each cluster 
hybridize with the mode of the Fermi surface at that site, 
leaving behind at low energies a chain of clusters with $N-1$ modes.
When $g/J$ is too small (at fixed $N$), the 
hybridization is unable to mix the levels
of the clusters, 
and in the infrared, the Fermi surface decouples; 
this phase is labelled FS $\times$ SYK.  
If $t$ is the smallest energy scale, we get 
decoupled clusters.
}
\end{figure}

Most of our work applies in any number of spatial dimensions, 
and only the discussion of \S\ref{sec:numerical-results} is specific to one dimension.
To emphasize this we include a diagram of the model in two dimensions
in Fig.~\ref{fig-2d}.
In Fig.~\ref{fig:phase-diagram} we sketch our picture of the phase diagram of the model
in the space of couplings $J/t, g/t$ studied here.

\section{Large-$N$ analysis}

\label{ref:large-N}

\subsection{SYK review}

We will use the complex fermion avatar of the SYK model described in \cite{Sachdev:2015efa}, and here we provide a brief description of its relevant known properties. 
The degrees of freedom are a set of canonical 
fermions $\rchi_i$
($ \{ \rchi_i, \rchi_j^\dagger \} = \delta_{ij} , \{ \rchi_i, \rchi_j\} = 0, i = 1..N$)
governed by the Hamiltonian \eqref{eq:HSYK}.
The object of interest to us is the disorder averaged fermion green function $\sG(\tau-\tau') = \overline{\vev{\rchi^\dagger_a(\tau) \rchi_a(\tau')}}$. This quantity can be calculated diagramatically by noticing that the only diagrams which survive disorder averaging are the ones in which interaction vertices can be grouped into pairs with identical indices. 

The diagrams contributing at leading order in $1/N$ are those in which the vertices are paired as locally as possible starting from the interior of the diagram moving outwards: this leads to the series of so-called melon diagrams.
This says that the one-particle irreducible part of $\sG$, the self energy $\Sigma$,
is itself a product of Green's functions: $\Sigma(\tau) =J^2 \mathcal \sG^2(\tau) \mathcal \sG(-\tau)$. Since there are no quadratic terms in $H$, the free propagator is $\mathcal \sG_0 = (\ii \omega)^{-1}$ in frequency space.

At small frequencies and strong coupling, the free $(\ii \omega)^{-1}$ part of Schwinger-Dyson equation for the fermion propagator $\mathcal \sG^{-1}(\omega) = (\ii \omega)^{-1} - \Sigma(\omega) $ becomes negligible compared to the self energy, resulting in 
the following closed set of integral equations for the Green's function:
$$ \int ds\; \mathcal \sG(\tau -s)\Sigma(s) \approx -\delta(\tau), \quad \Sigma(\tau) = J^2 \sG(\tau)^2 \sG(-\tau) = \raisebox{-0.5\height}{
\begin{fmffile}{black-melon}
  \begin{fmfgraph*}(80,80)
  \fmfleft{i}
  \fmfright{o}
  \fmf{fermion}{i,m1,m2,o}
  \fmf{phantom}{i,m1}
  \fmf{phantom}{m2,o}
  \fmf{fermion,left=0.6,tension=0}{m1,m2,m1}
  \fmf{dashes,left=1.3,tension=0}{m1,m2}
\end{fmfgraph*}
\end{fmffile}}.$$
These equations allow a power law solution for the Green's funciton; here we quote the result from \cite{Sachdev:2015efa}:
$$\mathcal G_{SYK}(\omega) = -\ii \left(\frac{\pi}{J^2}\right)^{1/4} \sqrt{\frac{2\beta}{\pi}} \frac{\Gamma(\tfrac{1}{4} + \tfrac{\beta \omega}{2\pi})}{\Gamma(\tfrac{3}{4} + \tfrac{\beta \omega}{2\pi})} ,~~ \mathcal G_{SYK} (\omega) = -\ii \left(\frac{\pi}{J^2}\right)^{1/4} \frac{\sgn \omega}{\sqrt{ |\omega|}}$$
at finite and zero temperature respectively, and at half filling. Away from half filling, the Green's function has a phase which we discuss in section \ref{sec:qto2}.
Most significantly, we note that the mass dimension of the SYK field is 
$\Delta(\rchi(\tau)) = - {1\over 4 }$.

We can consider generalizing the $(q=4)$-fermion interactions 
of $H_{SYK}$ to more general powers:
$ H(\rchi) = J_{i_1 \cdots i_q}  \rchi_{i_1}^\dagger \cdots \rchi_{i_q} $.
Redoing the above analysis gives 
$ \nu(q) = { 2 - q \over 2 q} $ and mass dimension $\Delta_q(\rchi(\tau)) = -\tfrac{1}{q}$.
We will take advantage of this parameter in \S\ref{sec:qto2}.

It is also possible to define - and consider coupling to - the 
{\it bath field} 
$$ \tilde \rchi_i  \equiv \sum_{jkl} J_{ijkl} \rchi_j^\dagger \rchi_k^\nd \rchi_l^\nd $$ 
which is the object multiplying $ \rchi_i$ in $H_{SYK}$.
The bath field has correlator and scaling dimension
$$ \vev{ \tilde \rchi^\dagger(\omega) \tilde \rchi^\nd(\omega)} \propto ( \ii \omega)^{+\half} ,~~~ \Delta(\tilde \rchi(t)) =  {3\over 4 }. $$
For general $q$, these are modified to
$\vev{ \tilde \rchi^\dagger(\omega) \tilde \rchi^\nd(\omega)} \propto ( \ii \omega)^{q-2 \over q}$ and $ \Delta(\tilde \rchi(t)) =  {q - 1 \over q }$.

\subsection{Using SYK clusters to kill the quasiparticles and take their momentum}
\label{sec:syk-kill}
The system we will study for the rest of the paper has
$ H = H_{FS}+H_{SYK}+H_{\text{int}}$ with 
$$ 
H_{FS} = \sum_{\vev{xy}} t \psi_x^\dagger \psi_y^\nd + h.c. 
= \int d^d k \eps(k) \psi_k^\dagger \psi_k^\nd
,~~
 H_\text{int} = \sum_{x,i} g^\nd_{ix} \psi^\dagger_x \rchi^\nd_{xi} + \textrm{h.c.}$$
We denote respectively the unperturbed and full $\vev{\psi\psi}$ propagators with a thin and thick black line.  
The $\vev{\chi\chi}$ propagator, denoted by a red line, includes the full series of melon diagrams. Disorder contractions are drawn as a dashed line. 
$$\raisebox{-0.5\height}{
\begin{fmffile}{black-line-arrow}
  \begin{fmfgraph*}(40,20)
  \fmfleft{i}
  \fmfright{o}
  \fmf{fermion}{i,o}
\end{fmfgraph*}
\end{fmffile}} = {1\over \omega - v_Fk_\perp},
~~ \raisebox{-0.5\height}{
\begin{fmffile}{red-line-arrow}
  \begin{fmfgraph*}(40,20)
  \fmfleft{i}
  \fmfright{o}
  \fmf{fermion,fore=red}{i,o}
\end{fmfgraph*}
\end{fmffile}}  =  \vev{\rchi^\dagger_x \rchi_y^\nd}, ~~
\raisebox{-0.5\height}{
\begin{fmffile}{disorder-contraction}
  \begin{fmfgraph*}(40,20)
  \fmfleft{i}
  \fmfright{o}
  \fmf{phantom}{i,o}
  \fmf{dashes,left=0.6,tension=0}{i,o}
\end{fmfgraph*}
\end{fmffile}}  = \text{disorder contraction} $$
The itinerant-fermion Green's function is given by a series of alternating $\psi$ and $\chi$ propagators. The only choice to make is how to contract the various interaction vertices in doing the Gaussian disorder averages. 
Any pattern other than the one shown below constrains an index sum over SYK flavors and is therefore suppressed by powers of $1/N$:
 $$\raisebox{-0.5\height}{\begin{fmffile}{black-full}
  \begin{fmfgraph*}(40,20)
  \fmfleft{i}
  \fmfright{o}
  \fmf{heavy,fermion,back=black}{i,o}
\end{fmfgraph*}
\end{fmffile}} 
= 
\raisebox{-0.5\height}{\begin{fmffile}{black-line-arrow}
  \begin{fmfgraph*}(40,20)
  \fmfleft{i}
  \fmfright{o}
  \fmf{fermion}{i,o}
\end{fmfgraph*}
\end{fmffile}}
+ 
\raisebox{-0.5\height}{\begin{fmffile}{ip1}
  \begin{fmfgraph*}(80,20)
  \fmfleft{i}
  \fmfright{o}
  \fmf{fermion}{i,m1}
  \fmf{fermion,fore=red}{m1,m2}
  \fmf{fermion}{m2,o}
  \fmf{dashes,left=0.7,tension=0}{m1,m2}
\end{fmfgraph*}
\end{fmffile}}
+
\raisebox{-0.5\height}{\begin{fmffile}{ip2}
  \begin{fmfgraph*}(140,20)
  \fmfleft{i}
  \fmfright{o}
  \fmf{fermion}{i,a}
  \fmf{fermion,fore=red}{a,b}
  \fmf{dashes,left=0.7,tension=0}{a,b}
  \fmf{fermion}{b,c}
  \fmf{fermion,fore=red}{c,d}
  \fmf{dashes,left=0.7,tension=0}{c,d}
  \fmf{fermion}{d,o}
\end{fmfgraph*}
\end{fmffile}} + \ldots$$

\begin{figure}[h] \begin{center}
 \includegraphics[width=0.5\textwidth]{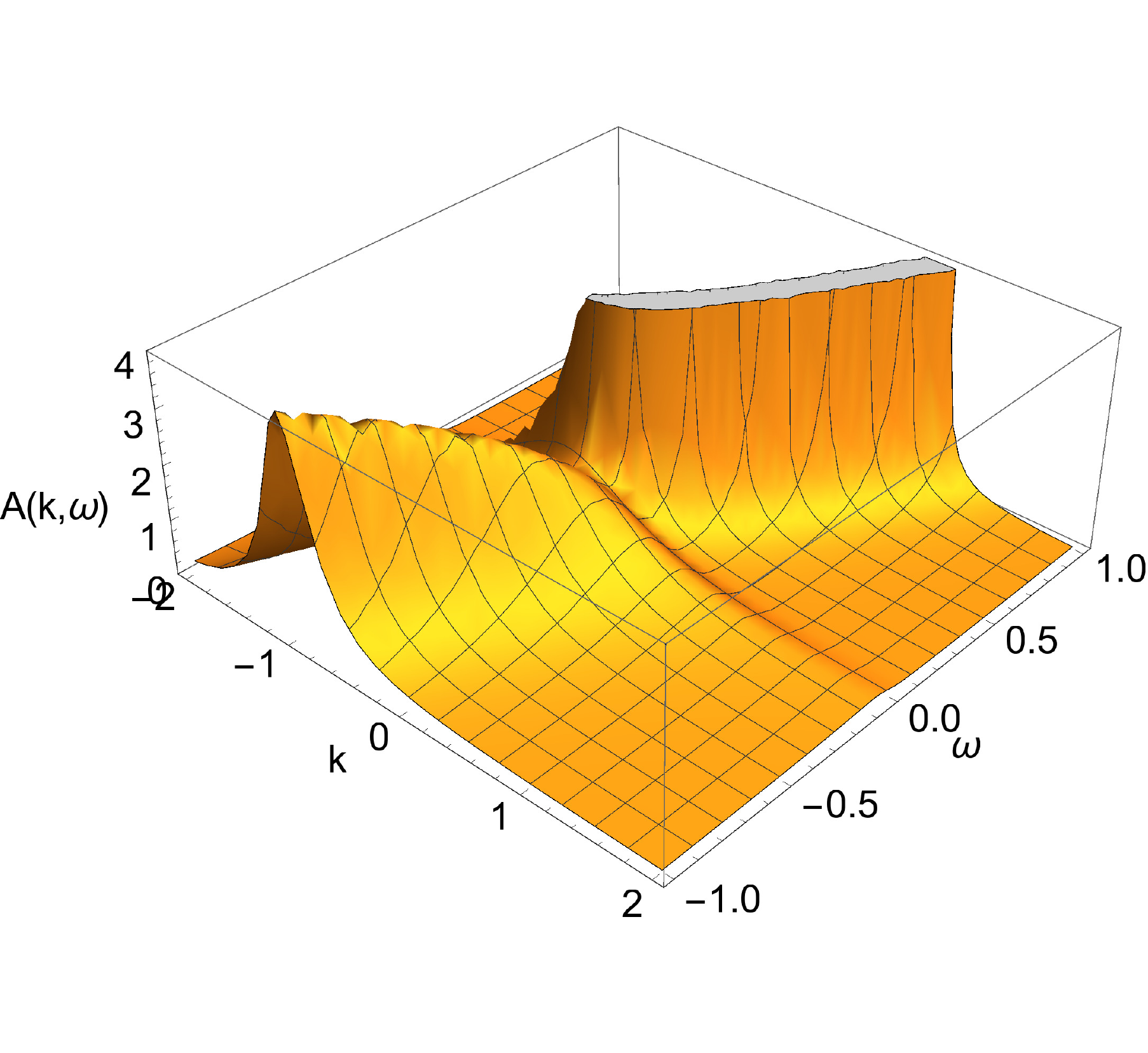}
\end{center}
  \caption{The self-energy diverges at $\omega=0$, leading
  to a {\it zero} of the Green's function, and of the spectral density $A(k,\omega)$, exactly at the Fermi surface. 
  \label{fig:diverging-self-energy}}
\end{figure}

Thus, the $\psi$ self-energy is $ \Sigma(\omega,k) = g^2 \redt{\mathcal G(\omega)}$
 (just as in the holographic model).  
 If we are interested in low energy physics near the Fermi surface, the SYK clusters are in the conformal limit and the Green's function behaves as
$$ G_\psi(\omega, k) \buildrel{\text{small $\omega$}}\over{=} 
 {1\over  \omega - v_F k_\perp - g^2 \redt{\mathcal G_{SYK}(\omega) }} $$
 This is of the same form as 
 found in the charged black hole calculation.
 In that context, various $\nu$ arise, but all have $ \nu \geq 0$.
In contrast, our model has $ \nu = - {1\over 4 } $, that is,
 $ \redt{\mathcal G(\omega)}  \sim \omega^{- {1\over 2 }}$.
This self-energy is not only non-analytic, but also {\it infinite} at 
$\omega \to 0$.  As a consequence, the 
Green's function {\it vanishes} at the Fermi surface.  
The spectral density $A(k,\omega) = {1\over \pi } \Im G(k,\omega)$ 
near the Fermi surface
is illustrated in Fig.~\ref{fig:diverging-self-energy}.
For general $q$, the exponent is $2\nu = {2\over q} - 1$, still negative
for all $q>2$.

Coupling to the bath field $\tilde\rchi$ would seem to give a more-familiar positive 
value of $ \nu = +{1\over 4 }$.
For general $q$, it would give $ \nu = { q - 2 \over 2 q } $, 
which approaches a marginal fermi liquid as $ q \to \infty$. 
We will see below, however, that this is a place where the $N \to \infty$ and 
low energy limits do not commute.

The conductivity from the itinerant fermions can be calculated using the Kubo formula. 
In the large $N$ limit, 
the transport analysis of \cite{Faulkner:2010da, Faulkner:2013bna}
(and related semi-holographic analyses \cite{Mukhopadhyay:2013dqa, Doucot:2017bdm})
can largely be carried over.  
The temperature dependence of the DC conductivity 
is a power-law determined by the localized-fermion 
Green's function exponent as $\sigma \propto T^{-2\nu}$. 
In particular, coupling to the SYK fermion and coupling to the bath field yield,
at large $N$, a resistance which is proportional to $T^{-1/2}$ and $T^{1/2}$ respectively. 

The divergence of the resistance as $T \to 0$ is related to the fact that the hybridization coupling is a relevant perturbation; this is a similar 
phenomenon to the resistance minimum
in the Kondo problem \cite{affleck2008impurity}.
In the Kondo case, the interaction is only marginally relevant, and hence
the resistance minimum occurs at an exponentially low scale;
here, for $q >2$ the interaction is relevant by a finite amount.
In the limit $ q \to 2$, 
the interaction becomes marginal,
suppressing the temperature at which the resistance rise sets in.
We study this limit further in \S\ref{sec:qto2}.

{\bf Does the Fermi surface delocalize the clusters?}

Contributions to the cluster fermion Green's function $\sG$ are again of the form
$$ \raisebox{-0.5\height}{\begin{fmffile}{red-line-arrow}
  \begin{fmfgraph*}(40,20)
  \fmfleft{i}
  \fmfright{o}
  \fmf{fermion,fore=red}{i,o}
\end{fmfgraph*}
\end{fmffile}}
+ 
\raisebox{-0.5\height}{\begin{fmffile}{lp1}
  \begin{fmfgraph*}(80,20)
  \fmfleft{i}
  \fmfright{o}
  \fmf{fermion,fore=red}{i,m1}
  \fmf{fermion}{m1,m2}
  \fmf{fermion,fore=red}{m2,o}
  \fmf{dashes,left=0.7,tension=0}{m1,m2}
\end{fmfgraph*}
\end{fmffile}}
+
\raisebox{-0.5\height}{\begin{fmffile}{lp2}
  \begin{fmfgraph*}(120,20)
  \fmfleft{i}
  \fmfright{o}
  \fmf{fermion,fore=red}{i,a}
  \fmf{fermion}{a,b}
  \fmf{dashes,left=0.7,tension=0}{a,b}
  \fmf{fermion,fore=red}{b,c}
  \fmf{fermion}{c,d}
  \fmf{dashes,left=0.7,tension=0}{c,d}
  \fmf{fermion,fore=red}{d,o}
\end{fmfgraph*}
\end{fmffile}}
+ 
\raisebox{-0.5\height}{\begin{fmffile}{lp3}
  \begin{fmfgraph*}(120,20)
  \fmfleft{i}
  \fmfright{o}
  \fmf{fermion,fore=red}{i,a}
  \fmf{fermion}{a,b}
  \fmf{dashes,left=0.5,tension=0}{a,d}
  \fmf{fermion,fore=red}{b,c}
  \fmf{fermion}{c,d}
  \fmf{dashes,left=0.7,tension=0}{b,c}
  \fmf{fermion,fore=red}{d,o}
\end{fmfgraph*}
\end{fmffile}}
+ 
\ldots
$$
where the only decision to be made is the manner of disorder contraction. 
Here is a place where the randomness of the hybridization
couplings $g_{ix}$ is crucial:
the processes by which $\sG_{xy}$ would develop off-diagonal terms
vanish by the disorder average over $g_{ix}$.  
The cluster fermions therefore stay localized, on average 
(however $ \overline{ \sG_{xy}  \sG_{xy} } $ will not be zero).

\begin{wrapfigure}{r}{0.4\textwidth}
\vspace{-25pt}
  $$
  \begin{fmffile}{turtle1}
  \begin{fmfgraph*}(120,20)
  \fmfleft{i}
  \fmfright{o}
  \fmf{fermion,fore=red}{i,m1}
  \fmf{fermion}{m1,m2}
  \fmf{fermion,fore=red}{m2,o}
  \fmf{dashes,left=0.7,tension=0}{m1,m2}
\end{fmfgraph*}
\end{fmffile} + ~~~~
  $$
    $$
  \begin{fmffile}{turtle2}
  \begin{fmfgraph*}(120,20)
  \fmfleft{i}
  \fmfright{o}
  \fmf{fermion,fore=red}{i,a}
  \fmf{fermion}{a,b}
  \fmf{fermion,fore=red}{b,c}
  \fmf{fermion}{c,d}
  \fmf{fermion,fore=red}{d,o}
  \fmf{dashes,left=0.4,tension=0}{a,d}
  \fmf{dashes,left=0.7,tension=0}{b,c}
\end{fmfgraph*}
\end{fmffile}+~~~~
  $$
    $$
  \begin{fmffile}{turtle3}
  \begin{fmfgraph*}(120,20)
  \fmfleft{i}
  \fmfright{o}
  \fmf{fermion,fore=red}{i,a}
  \fmf{fermion}{a,b}
  \fmf{fermion,fore=red}{b,c}
  \fmf{fermion}{c,d}
  \fmf{fermion,fore=red}{d,e}
  \fmf{fermion}{e,f}
  \fmf{fermion,fore=red}{f,o}
  \fmf{dashes,left=0.4,tension=0}{a,f}
  \fmf{dashes,left=0.7,tension=0}{b,c}
  \fmf{dashes,left=0.7,tension=0}{d,e}
\end{fmfgraph*}
\end{fmffile}+~~~~
  $$
      $$
  \begin{fmffile}{turtle4}
  \begin{fmfgraph*}(120,40)
  \fmfleft{i}
  \fmfright{o}
  \fmf{fermion,fore=red}{i,a}
  \fmf{fermion}{a,b}
  \fmf{fermion,fore=red}{b,c}
  \fmf{fermion}{c,d}
  \fmf{fermion,fore=red}{d,e}
  \fmf{fermion}{e,f}
  \fmf{fermion,fore=red}{f,g}
    \fmf{fermion}{g,h}
  \fmf{fermion,fore=red}{h,o}
  \fmf{dashes,left=0.4,tension=0}{a,h}
  \fmf{dashes,left=0.7,tension=0}{b,c}
  \fmf{dashes,left=0.7,tension=0}{d,e}
  \fmf{dashes,left=0.7,tension=0}{f,g}
\end{fmfgraph*}
\end{fmffile}+ \cdots
  $$
  \caption{The corrections to the localized-fermion propagator $\sG$ at order $1/N$. \label{fig:turtles}}
\end{wrapfigure}
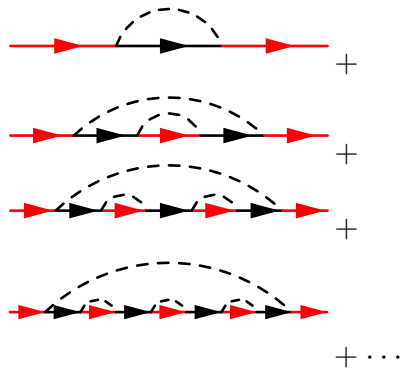
Futhermore, the onsite corrections to the SYK Green's function are small; they are of order $1/N$. 
The leading order correction is obtained by 
summing the `turtle' diagrams in Fig.~\ref{fig:turtles}. Taken together, this series of diagrams combines into the object $\tfrac{g^2}{N} \sG_0^2(\omega) \int \dbar^d k \; G(k\omega)$, as we show in Appendix \ref{app:turtles}. Thus there is the possibility that the SYK-ness of the cluster fermion will be disrupted at parametrically low energies. In Appendix 
\ref{app:turtles}, we show that this correction in fact does not modify the leading low frequency behavior, even at frequencies small compared to $1/N$.

Now consider the effects of $\delta \sG$ on the itinerant propagator.  
The leading-in-$N$ self-energy $\sG(\omega)$ itself diverges like $ \omega^{-1/2}$ at low frequency.
It therefore dominates over $\delta \sG$, 
which vanishes at asymptotically small $\omega$, as we show in Appendix \ref{app:turtles}.

\subsection{Replica analysis}

\label{sec:replicas}

The leading-order diagrammatic calculation above can be reproduced by a replica calculation. It suffices to consider a single cluster impurity. The replicated action before any disorder averaging is 
\begin{equation*}\begin{split} S[\psi,c] = 
\sum_a\int d\tau  \sum_{i} \bar c_{ia\tau}\partial_\tau c_{ia\tau}
+ \bar f_{xa\tau} \( \partial_\tau  - \xi(\partial_x) \) f_{xa\tau} 
 \\ + \sum_{ijkl} J_{ijkl} \bar c_{ia\tau} \bar c_{ja\tau} c_{ka\tau} c_{la\tau} 
+ \sum_i g_{i} \bar c_{ia\tau} f_{a\tau}(r_{imp}) + h.c. 
\end{split}
\end{equation*}
Here $\xi(k)$ is the band dispersion.  
In what follows, we will occasionally drop the time arguments for compactness of writing. In that case the argument $\tau$ is always accompanies the replica index $a$ and $\tau'$ with $b$. Averaging over $g$s with a gaussian weight of width $g$ produces a term 
$$
\CI\equiv \exp{\left( \frac{g^2}{2N}  \int d\tau d\tau' \bar f_{a}(r_{imp},\tau)f_{b }(r_{imp},\tau') \sum_{i} c_{a\tau  i } \bar c_{b\tau' i }  \right)}.
$$
This is decoupled with two 
hermitian
Hubbard Stratonovich (HS) fields $\rho_{ab}(\tau,\tau')$ and $\sigma_{ab}(\tau,\tau')$. 
By `hermitian', we mean $ \rho_{ab}(\tau, \tau') = \rho^\star_{ba}(\tau', \tau)$.
$$\CI =  \int D \rho_{ab} D \sigma_{ab}\; \exp\left[ -\frac{1}{2}\int d\tau d\tau' \; N \sum_{ab} \rho_{ab}^2(\tau,\tau') + \sigma_{ab}^2(\tau,\tau')   +\frac{1}{2}\int d\tau d\tau' g \left( \sum_{ab} \rho^\nd_{ab} F^-_{ab} + \ii \sigma^\nd_{ab} F^+_{ab} \right) \right] $$
where $F^{\pm}_{ab} = \bar \psi_a(\tau) \psi_b(\tau') \pm  \sum_i \bar c_{ia}(\tau) c_{ib}(\tau')$.

Introducing two sets of Hubbard-Stratonovich fields following Bray-Moore \cite{bray1980replica} and Sachdev \cite{Sachdev:2015efa}, we can factorize the contribution from the average over $J$ as 
\be
\begin{split}
 &  
 \int \prod_{ijkl} dJ_{ijkl} ~e^{ - N^3 { J_{ijkl}^2\over  J^2}}
 e^{ \int dt J_{ijkl} \bar c_i c_j \bar c_k c_l } = 
 e^{{J^2 \over 4N}\sum_{ab} \int dt \int dt' | \sum_i \bar c_{iat} c_{ibt'} |^4 } 
\\ =&  \int [dQ dP] \exp\left({ \int d\tau d\tau' \sum_{ab} \(  -{ N \over 4 J^2 } Q_{ab}(\tau,\tau')^2 - {N\over 2} Q_{ab} |P_{ab}(\tau,\tau')|^2 + Q^\nd_{ab}P^\nd_{ba} \sum_i \bar c_{ia} c_{ib} \) }\right)
\end{split}
\label{eq:HSx2}
\ee
where $Q$ and $P$ are real and complex symmetric and hermitian fields, respectively. Dropping the ``site'' index on the cluster fermions, the replicated disorder averaged action takes the form $\sum_{ab} S_0[\psi] + N S_1[c] + N S_2(\rho,\sigma,Q,P) $ with 

\begin{eqnarray*}
S_0[\psi] &=&  \int d\tau d\tau' d^dx \; \bar \psi_{ax\tau} \left(\delta_{ab} \partial_\tau - \delta_{ab} \xi(\partial_x) - \frac{g}{2} \delta^d(x-r_{imp}) (\rho_{ab} + \ii \sigma_{ab}) \right) \psi_{bx\tau'}  ,\\
S_1[c] &=&  \int d\tau d\tau' \;  \bar c_{a \tau}\left( \delta_{ab}\partial_\tau  + \frac{g}{2} ( \rho_{ab} - \ii \sigma_{ab}) -  Q_{ab}P_{ba} \right)c_{b\tau'},\\
S_2(\rho,\sigma,Q,P) &=&  \frac{1}{2} \int d\tau d\tau'\left( \rho_{ab}^2 + \sigma_{ab}^2 +
 \frac{1}{2 J^2} Q_{ab}^2 +  Q_{ab} |P_{ab}|^2 \right).
\end{eqnarray*}
The saddle point equations for $Q,P,\rho$, and $\sigma$ resulting from this effective action
give us the standard SYK saddle point results
$$P_{ab} = \vev{\bar c_{a\tau} c_{b\tau'}} ,~~ Q_{ab} = J^2 |P_{ab}|^2 $$
supplemented by two additional relations for the fields $\rho \pm \ii \sigma$.
$$ \rho + \ii \sigma = -g \vev{\bar c_{a\tau} c_{b\tau'} },~~ \rho - \ii \sigma = \frac{g}{N} \vev{\bar \psi_{a\tau,x=0} \psi_{b\tau,x=0}}.$$

Upon integrating out $\psi$ and $c$ degrees of freedom, 
and assuming no replica symmetry breaking (and setting the position of the cluster at the origin $r_{imp} = 0$) one finds the effective action
$$ S_{eff} = S_2(\rho,\sigma,Q,P) - \ln \det \left( \partial_\tau - QP + \frac{g}{2}(\rho - \ii \sigma) \right) - \ln \det \left(\partial_\tau - \xi(\partial_x) -\frac{g}{2} \delta^d(x) (\rho + \ii \sigma) \right).$$
We can identify
$$ \Sigma_{SYK} =   J^2 \mathcal G |\mathcal G|^2  +\frac{g^2}{N}  G_{\psi}(x,x),~~ \Sigma_{\psi}  = g^2 \mathcal G_{syk}$$
which reproduces the previous result.
In fact the replica analysis goes a step beyond the analysis of the previous section: it sums the series of corrections to the SYK propagator in powers of $g^2/N$, of which we only explicitly analyzed the first term. The $\delta(x)$ in these equations appears because we studied a single impurity, and yields a momentum-independent self energy upon Fourier transforming.

\section{Finite $N$} 
\label{sec:finite-N}

\subsection{Renormalization group analysis of impurity problem}

Consider a single SYK cluster coupled to the itinerant mode.
There is quite a bit of physics in this impurity problem, 
and it will be an extremely useful starting point.
As we noted in \S\ref{sec:replicas}, the large-$N$ analysis is basically identical.

{\bf Weak coupling.}  First consider the regime where $ g \ll t, J$.
In this case, the correlation length of $\psi$ is large 
compared to the lattice spacing, 
and we can treat the itinerant fermions in the continuum.
Following the  literature on the Kondo problem \cite{affleck2008impurity}, only the $s$-wave 
mode 
of the Fermi surface 
 $ \psi_0(k) \sim k \int d\hat\Omega \psi(\Omega k)$
couples. Linearizing the $s$ wave mode near the fermi surface with a bandwidth cutoff $\Lambda$ the Hamiltonian for the left/right moving fields $\psi_{L/R} = \int_{-\Lambda}^{\Lambda} dk\; e^{\pm \ii k r} \psi_0(k+k_f)$  is 
\cite{affleck2008impurity}
$$ H_{FS} = { v_F \over 2\pi} \int_0^\infty dr \( \psi_L^\dagger \partial_r \psi_L^\nd
- \psi_R^\dagger \partial_r \psi_R^\nd \)$$
This implies that the free fields under consideration have mass dimension $ [\psi_{L/R}] = \half. $ The scaling dimension of the SYK fields was determined in the low energy analysis in the previous section and found to be $1/4$ for the fermion field and $3/4$ for the bath field.  The perturbation we are considering are of the forms
$$ \Delta H = g \psi_L^\dagger(0) \rchi, 
~~~~~~~~~~\Delta \tilde H = \tilde g \psi_L^\dagger(0) \tilde \rchi. $$
The scaling dimension of the coupling constant $g$ determines whether the hybridization becomes more or less important at low energies. 
Demanding that the action is dimensionless, the coupling to the bath field has mass dimension
$- [ \int dt~\psi^\dagger \tilde \rchi ] =- \(  -1 + \half + {3 \over 4 }\) = - {1 \over 4 } $
and is therefore irrelevant.
The coupling to the fundamental field $\rchi$ has dimension $ - [ \int dt\; \psi^\dagger \rchi ] = - \( - 1 + \half + {1 \over 4 }\)= +{1 \over 4 } $,
and is therefore relevant.
Here again we depart from the holographic construction,
where $\sG \sim \omega^{2\nu}$ with positive $\nu$ -- according to the above analysis 
the construction studied here, only $\sG$ with negative $\nu$  can 
dominate the infrared physics.

{\bf Strong coupling.}  Now consider the regime where $ g \gg t, J$.
This is a highly-underscreened Anderson model.
At each site, the itinerant fermion $\psi(x)$
is coupled to a particular linear combination ${1\over N} \sum_i g_i \rchi_i(x) \equiv \tilde \rchi_N(x)$ 
of the SYK fermions at site $x$.  
Take linear combinations of the $\rchi_i$ to orthogonalize the first $N-1$ with $ \tilde \rchi_N$.
Then in the limit where $|g \gg J|$ {(where $g$ is the average of the $g_i$)},
we can simply neglect the four-fermion interactions involving $\tilde c_N$
and the result of the hybridization is simply
to pair up $\psi(x)$ and $ \tilde \rchi_N(x)$ at each site,
leaving behind at low energy only $N-1$ decoupled SYK clusters.
This can be called a reverse Kondo phase: 
whereas the Kondo effect describes the absorption of
an impurity into the Fermi sea of conduction electrons,
here the situation is reversed: the impurities absorb the conduction electrons!

\begin{figure}[h] \begin{center}
\parfig{.25}{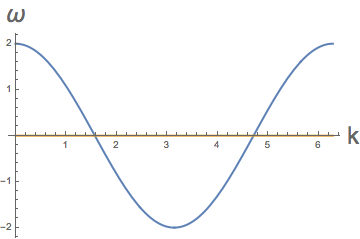} $\to$
\parfig{.25}{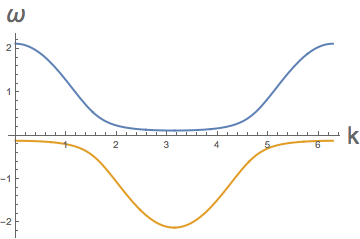}
\end{center}
\caption{
When $g \gg J$, we can neglect the SYK interactions,
and our problem becomes quadratic.  
Hybridizing a localized fermion (flat band) 
with an itinerant fermion 
produces this bandstructure.
\label{fig:free-localized-fermion-hybridization}
}
\end{figure}

\subsection{Possibilities for the phase diagram}

Considerations of the topology of coupling space
constrain the possibilities 
for the low-energy behavior of our system.
Given that $g$ is a relevant perturbation
of the decoupled fixed point with $g=0$, 
and given that at large $g$, it produces
a mass gap, the possible RG flow
diagrams are as follows:
\begin{figure}[h] 
\begin{center}
1:~ \parfig{.25}{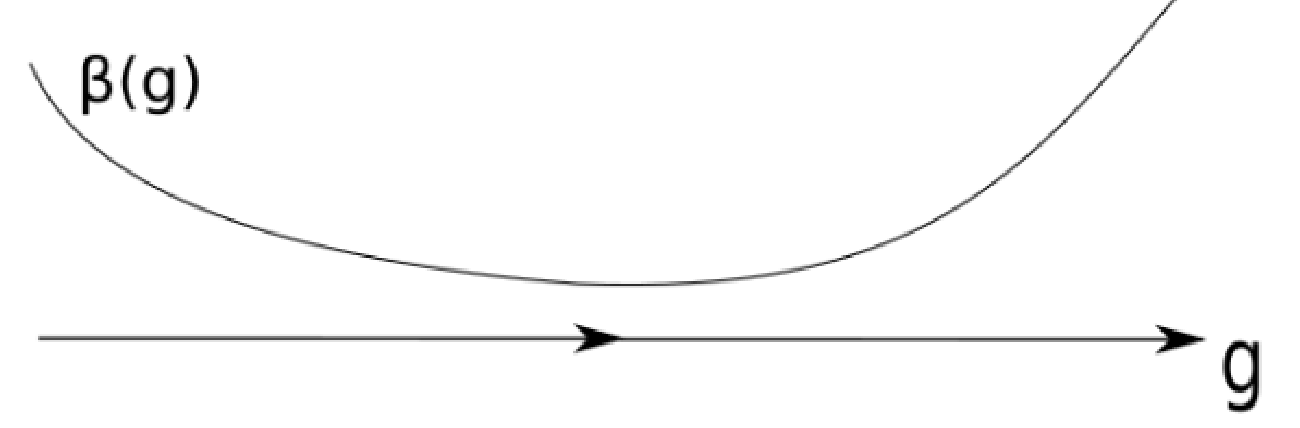}~2:
~\parfig{.25}{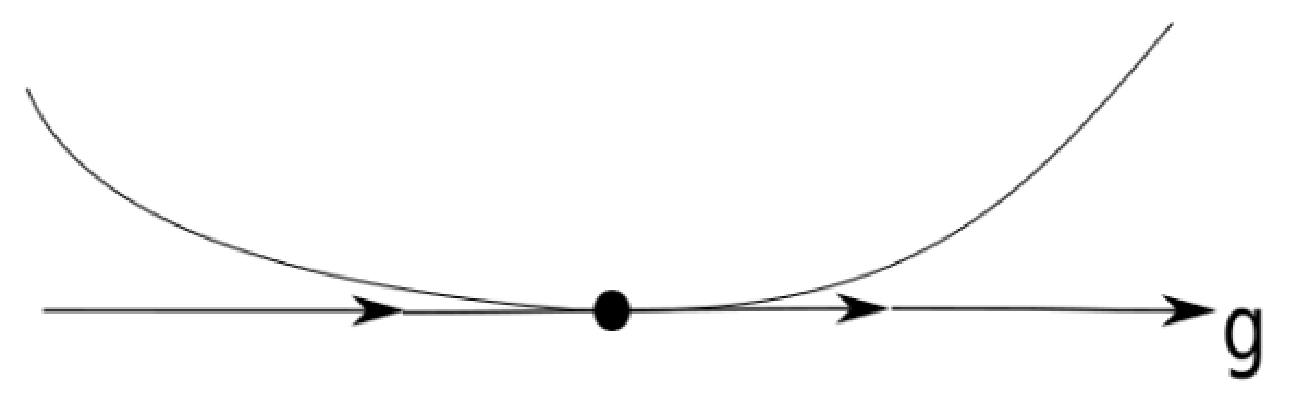}~3:~
\parfig{.25}{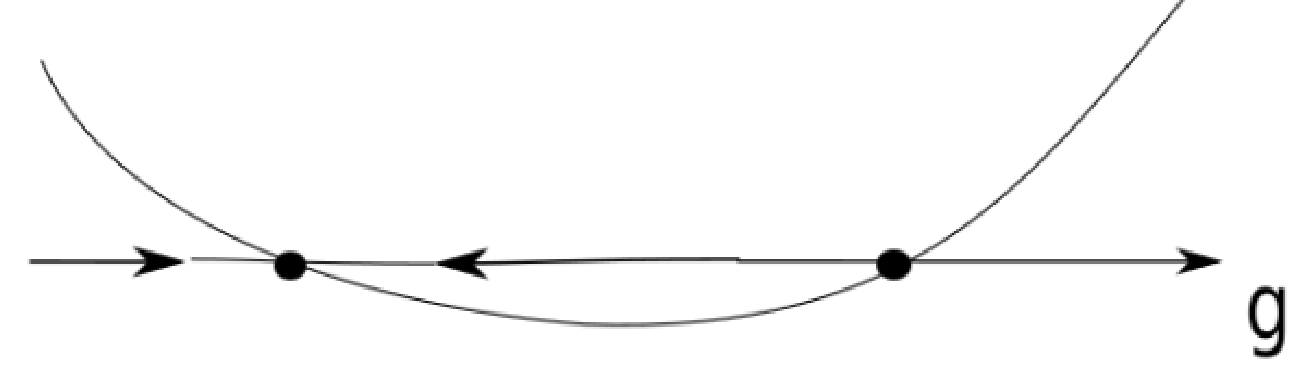}
\end{center}
\vskip-.1in
\caption{Possible behaviors 
of the beta function for $g$, 
given the known asymptotics.
Arrows point toward the infrared.
\label{fig:RG-scenarios}
}
\end{figure}

The middle case (2) is nongeneric\footnote{A well-known example where the beta function 
has a double zero is in the BCS phase diagram, 
where $ \beta(V) \propto V^2 + ...$.  Here the double zero
occurs at the free theory, and is therefore protected by dimensional analysis.}.
Therefore, if we find a fixed point, it is stable.
In the following section, we will study the half-chain entanglement
entropy.  
The above scenarios for the beta function 
would imply the following 
rough consequences for this quantity,
respectively:
\begin{figure}[h] \begin{center}
1:~\parfig{.4}{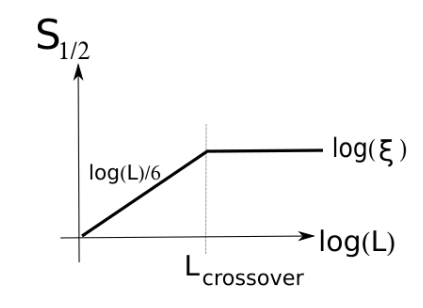}
~~
3:~\parfig{.4}{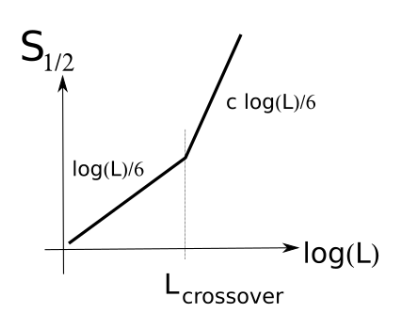}
\end{center}
\caption{Behavior of the half-chain entanglement 
entropy in scenarios 1 and 3 of Fig.~\ref{fig:RG-scenarios}.
\label{fig:RG-scenarios}
}
\end{figure}

The respective scenarios {\it would} imply these behaviors 
of the half-chain entropy {\it if} the system
were translation invariant.  Although 
there are examples of highly-disordered fixed points
which exhibit 
logarithmic area-law violation \cite{refael2004entanglement, refael2009criticality}, 
it is not clear whether this is inevitable.

We note that the behavior in scenario 3 does
not violate RG monotonicity \cite{zamolodchikov1986irreversibility} of 
the `central charge',
because the UV fixed point 
is tensored with decoupled, localized clusters
and is not a field theory.
More generally, in a system without Lorentz symmetry, $ \( { d  \over d \log L}\)^2 S_\half$ 
may be positive.
There are indeed known examples 
of disordered 
\cite{santachiara-2006}
and otherwise non-relativistic systems
\cite{swingle2014entanglement}
where the `central charge' (coefficient of $\log L$) 
increases towards the infrared.
Note further that 
the condition we are violating,
$ \( { d  \over d \log L}\)^2 S_\half < 0 $,
is a stronger condition than 
$ { d^2 \over d L^2 } S_\half < 0 $.  
In any case, the quantity $S_\half(L)$ does not obey a known convexity theorem
analogous to that of \cite{2014arXiv1405.1471G}, 
which applies instead to the entanglement entropy 
as a function of subsystem of size $ \ell$ 
of a {\it fixed} total system\footnote{Thanks to Tarun Grover for helpful discussions of these constraints
on the behavior of $S(\ell, L)$.}.

\section{Coupling a Fermi surface to SYK${}_{2.0001}$ clusters} 
\label{sec:qto2}

In the limit $ q \to 2$, the coupling $ \int \psi \chi^\dagger$  becomes 
marginal.
Therefore, in this limit, there is a hope that the NFL fixed point we're
after can be accessed perturbatively in $g$.  
Indeed, as we sketch here, this seems to be the case.

Consider the replicated and disorder-averaged euclidean partition function\footnote
{In many disordered systems, 
one must consider 
the RG evolution of the probability distribution for the disorder.
The renormalization group strategy pursued here, 
of studying the flow of the disorder-averaged action, 
{\it assumes} that the Gaussian disorder-distribution
for $g_{ix}$
is self-similar under an RG transformation --
we are allowing only its variance $g$ to evolve.  
} at $T=0$
$$ Z^n = \int [d\psi d\chi] e^{ - S_0 - g^2 a^{2-2\Delta(q)} 
 \int d\tau \int d\tau' 
 \sum_{x,i}
 \psi_{x}(\tau) \chi_{xi}(\tau)^\dagger 
\chi_{xi}(\tau')\psi_x(\tau')^\dagger } .$$
Replica indices accompany the time labels and are suppressed.  
Here $S_0$ is the action for the fixed point described by 
a Fermi surface
$$ \vev{ \psi^\dagger(\omega, k) \psi^\nd(\omega,k)}_{0} = { 1\over \ii \omega - v_F |k-k_F| } $$
times 
decoupled SYK${}_q$ clusters at each site {\it in their conformal limit}, 
$$ \vev{ \chi^\dagger(\tau) \chi^\nd(0) }_{0} = C(J) \text{sign}(\tau) |\tau|^{-2/q}. $$
Here $C(J) = C J^{-2/q}$
with
$C >0$ \cite{Sachdev:2015efa, Davison:2016ngz}.
The factor of $a^{2- 2\Delta(q) }$ 
(where $2 - 2\Delta(q) = {q-2 \over  q } $ is the 
scaling dimension of $\chi \psi^\dagger$)) has been pulled out of $g$ 
to make $g$ dimensionless.

We implement the RG as in \cite{cardy1996scaling}, by expanding
\be
\begin{split} 
Z^n = {} & Z^n_\star 
\( 1 - g^2 \vev{\int d\tau \int d\tau' \psi(\tau) \chi(\tau)^\dagger 
\chi(\tau')\psi(\tau')^\dagger }_{0} \right.
\\ + {} & g^4 
\left. \vev{\int d\tau \int d\tau' \psi(\tau) \chi(\tau)^\dagger 
\chi(\tau')\psi(\tau')^\dagger \int d\tau'' \int d\tau''' \psi(\tau'') \chi(\tau'')^\dagger 
\chi(\tau''')\psi(\tau''')^\dagger }_{0}  + \cdots 
\) .
\end{split}
\label{eq:conformal-perturbation-theory}
\ee
We wish to let $g = g(a)$ run with the UV cutoff $a$ in 
such a way as to cancel the dependence of $Z$ on $a$,
perturbatively in $g$.  
The cutoff dependence appears explicitly in the perturbation
term and implicitly in the need to regulate collisions of the integrations
$ |\tau - \tau ''| >  a $.  

The contractions in the 
$\CO(g^2)$  term 
produce corrections to the renormalized action of the form 
$$ \delta S  = g^2 \int d\tau  \( \psi^\dagger(\tau ) \psi^\nd(\tau) A + \chi^\dagger(\tau) \chi^\nd(\tau) B \)$$
where 
\bea B &=& \int_a d\tau \vev{ \psi^\dagger(\tau,x) \psi^\nd(0,x)}_{0} 
= \int_a d\tau \int \dbar \omega e^{ -  \ii \omega \tau} \int {\dbar^dp \over \ii \omega - v_F |p-k_F| } 
\cr &=& 
 \int_a d\tau \int \dbar^d p e^{ - v_F p_\perp\tau } 
\cr &\simeq &
\half \int_{-\beta}^\beta d\tau \underbrace{\Omega_{d-1}  \over (2\pi)^d}_{\equiv K_d} k_F^{d-1} 
\underbrace{\int d p_\perp  
e^{ - \ii  v_F p_\perp \tau } }_{= {1 \over v_F\tau }}
=
- {K_d k_F^{d-1} \over v_F} \( \int_{-\beta}^a + \int_a^\beta \){ d\tau \over \tau }  = 0
\label{eq:chemical-potential-correction}
\eea
and
$$ A = \int_a d\tau \vev{ \chi^\dagger(\tau) \chi^\nd(0)}_{0}  
= C(J) \int_a d\tau \text{sign}(\tau) |\tau|^{- 2/q}   = 0. $$
In \eqref{eq:chemical-potential-correction}, $\beta$ was introduced as an IR regulator\footnote{
Note that since zero is a bosonic matsubara frequency,
it is important that we integrate from $ - \beta $ to $\beta$ (and divide by two), 
rather than just $a$ to $\beta$. 
The latter would give $B \buildrel{?}\over{\sim} \log a T $.
Thanks to Aavishkar Patel for patient explanations of this point.
}.
$A$, were it nonzero, would be an innocuous correction to the $\psi$ chemical potential.  
Away from half-filled clusters, where $|\sG(\tau)| \neq |\sG(-\tau)|$, 
we find $ A \sim  a^{1 - 2/q} $.
Similarly, $B$ would be a correction to the chemical potential for $\chi$.  
In \cite{Sachdev:2015efa, Davison:2016ngz}, such a chemical potential is included in the analysis;
the phase of $\sG$ depends on it, but it is otherwise innocuous as well.

The interesting term for us 
is the (connected) contraction of the $g^4$ term 
which renormalizes $g^2$.   
This is 
\begin{fmffile}{beta-function}
\bea\delta g^2 &=& 
~~
- \half~ \parbox{0.15\textwidth}{\begin{fmfgraph}(60,40)
  \fmfleft{i}
  \fmfright{o}
    \fmf{fermion, fore=red,back=red, tension=0, right}{o,i}
  \fmf{fermion, fore=black,back=black, tension=0.01}{i,v1}
    \fmf{fermion, fore=black,back=black, tension=0.01}{v1,o}
     \fmfdot{o,i}
      \fmfblob{.5w}{v1}
\end{fmfgraph}
} 
\cr &=& 
(-1)^2 \half \int_a d\tau 
 \vev{ \chi^\dagger(\tau) \chi^\nd(0)}_{0} 
 \vev{ \psi^\dagger(\tau,x) \psi^\nd(0,x)}_{0}  + h.c. 
 \cr 
 && ~~~  ~~~~{}
 \cr
 &\simeq& \half C(J) {K_d k_F^{d-1} \over v_F} \int_a d\tau \tau^{- 2/q} \tau^{-1} 
 \simeq \half C(J) {K_d k_F^{d-1} \over v_F} a^{  - 2/q } ~.
\eea
\end{fmffile}
The minus sign in the first line is from the relative sign between
the $\CO(g^4)$ term and the $\CO(g^2)$ term in \eqref{eq:conformal-perturbation-theory}.
The minus sign in the second line is from the fermion loop --
we are contracting non-adjacent fermion operators.
Crucially, $C(J)$
is {\it positive}
for all values of the parameter $\theta$ (which is determined by the filling).

Therefore, 
$$ \beta_{g^2} \equiv { d \over d \log a} g^2 
= ( 2 -2 \Delta(q)) g^2 + \half C(J) {K_d k_F^{d-1} \over v_F} \( - {2 \over q } \) g^4 + \CO(g^6) . $$
Here 
$$  2 -2 \Delta(q)  = {q-2 \over q } = 1 - { 2 \over 2 + \eps}  =  \eps  + \CO(\eps^2) , ~~
- {2 \over q } = - { 2 \over 2 + \eps} = -1 + \eps + \CO(\eps^2)
.$$
Besides the trivial fixed point at $g=0$, this indicates a fixed point $ 0 = \beta_{g^2}(g= g_\star)$ 
at
$$ g_{\star}^2 = \frac{2 v_F  }{  C(J) K_d k_F^{d-1}  }  \eps + \CO(\eps^2),
$$
which is indeed at weak coupling, parametrically in $\eps$.
We note that it is also parametrically small in 
the area of the Fermi surface, $k_F^{d-1}$, 
suggesting that perhaps the physics at $q=4$ can be captured by this analysis.  
The fixed point depends on $J$ like 
$g_\star^2 \sim C(J)^{-1} \sim J^{2/q}.$

\section{Numerical analysis}
\label{sec:numerical-results}
We have attempted to perform some quantitative studies 
of the model considered in this paper, in the special case of a one-dimensional chain.
We use the standard technique for numerical studies of one dimensional systems, the density matrix renormalization group (DMRG) \cite{white1992density}. Specifically, we use a single site matrix product state sweeping algorithm \cite{schollwock2011density}. There are several factors which make it difficult to study this system numerically. 

The unusually large size of the local Hilbert space at each site (which is $2^{N_{syk} + 1}$, 
as opposed to $2$ for a spin 1/2 chain or $4$ for spinful fermions) means that the computational resources required at a given bond dimension are significantly larger than what is needed for studying spin chains. 
Furthermore, as is the case in most studies of systems with quenched disorder, we are interested in correlation functions averaged over many disorder realizations. Therefore, at each set of coupling constants, we must perform enough trials to achieve convergence. In some cases the number of trials required is relatively small ($\sim 50$) and in other cases it is larger $(\sim 500)$.

We use two different methods for our DMRG study. One is a completely standard MPS based DMRG sweeping algorithm in which we take $N_{syk} = 6$ on each site. This number is not very large, 
but is perhaps comparable to the numbers one
might hope for in material realizations of such a model.

The other method, 
which we'll refer to as the truncation method, begins with $N_{syk} = 12$ on each site. The size of the local Hilbert space here is too large to work with in our DMRG algorithm, so we form an isometry which projects the Hamiltonian into the subspace spanned by the $128=64*2$ lowest energy eigenstates. That is, we exactly diagonalize $\HH_{syk}$ with 12 modes in the presence of one extra fermionic mode which the Hamiltonian doesn't act on. So we truncate a Hilbert space of the form $2^{12} \otimes 2 \to 64_{syk} \otimes 2$. The entire Hamiltonian, as well as the hybridization and hopping terms, are projected into this truncated space.

The idea behind the truncation approach is that the properties 
of interest (in particular, the singular self-energy)
arise due to the special low energy physics of the SYK cluster. The expressions given for the Green's functions of the large $N$ theory considered in section \ref{sec:syk-kill} were all valid at low energies and at momenta near the Fermi surface. 
The relevant energy scales to compare are the hybridization coupling $g$ and the bandwidth $D$ of the states
that are retained. This is found to be  $D \sim 0.26 J$ at $N_{syk} = 12$. 

To help map out the phase diagram,
one of the most convenient and easily accessible quantities we can measure is the entanglement entropy 
of subregions of the chain (EE). In particular, (a review is \cite{Calabrese:2009qy}) a one dimensional conformal field theory (CFT) in the thermodynamic limit has an entanglement entropy which grows with the size $L$ of the subregion as $\tfrac{c}{3} \log L$, where $c$ is the central charge of the CFT\footnote{For simplicity, 
we assume a non-chiral spectrum.}. Similarly, for a CFT on a space of length $L$,
the half-chain entanglement entropy scales with the system size as $\tfrac{c}{6} \log L$. Thus, measuring the growth of the half-chain entanglement entropy with the system size allows us to access some universal information about the phase
and its low-energy excitations, from just the groundstate wavefunction.
We note that the emergence of Lorentz symmetry, 
much less conformal symmetry, is unlikely in our disordered system, 
so the measured behavior of the entanglement entropy is a proxy
for the number of low-energy degrees of freedom.

Considering fixed $J$, we know the behaviour of the half-chain EE at both small $g$ and very large $g$. At zero $g$,
the SYK clusters are decoupled from the free fermion chain. The latter
is responsible for all of the spatial entanglement, and has $c = 1$ for spinless fermions.  That is what we observe from the slope of the half chain EE. At large $g$, the hybridization term dominates and we expect the itinerant fermions to bind into a local singlet. This phase has a finite correlation length which becomes very small at large $g$.
Hence the EE satisfies an area law and $c = 0$. We observe this behaviour in our simulations.

As $g$ increases from zero, there are two possibilities, as we discussed in \S\ref{sec:finite-N}. Although finite size effects are hard to overcome in our particular model, measuring the slope of the half chain EE at different values of $g$ provides some evidence for either scenario 1 or 3 above. 

\begin{figure}[h] \begin{center}
\includegraphics[width=.7\textwidth]{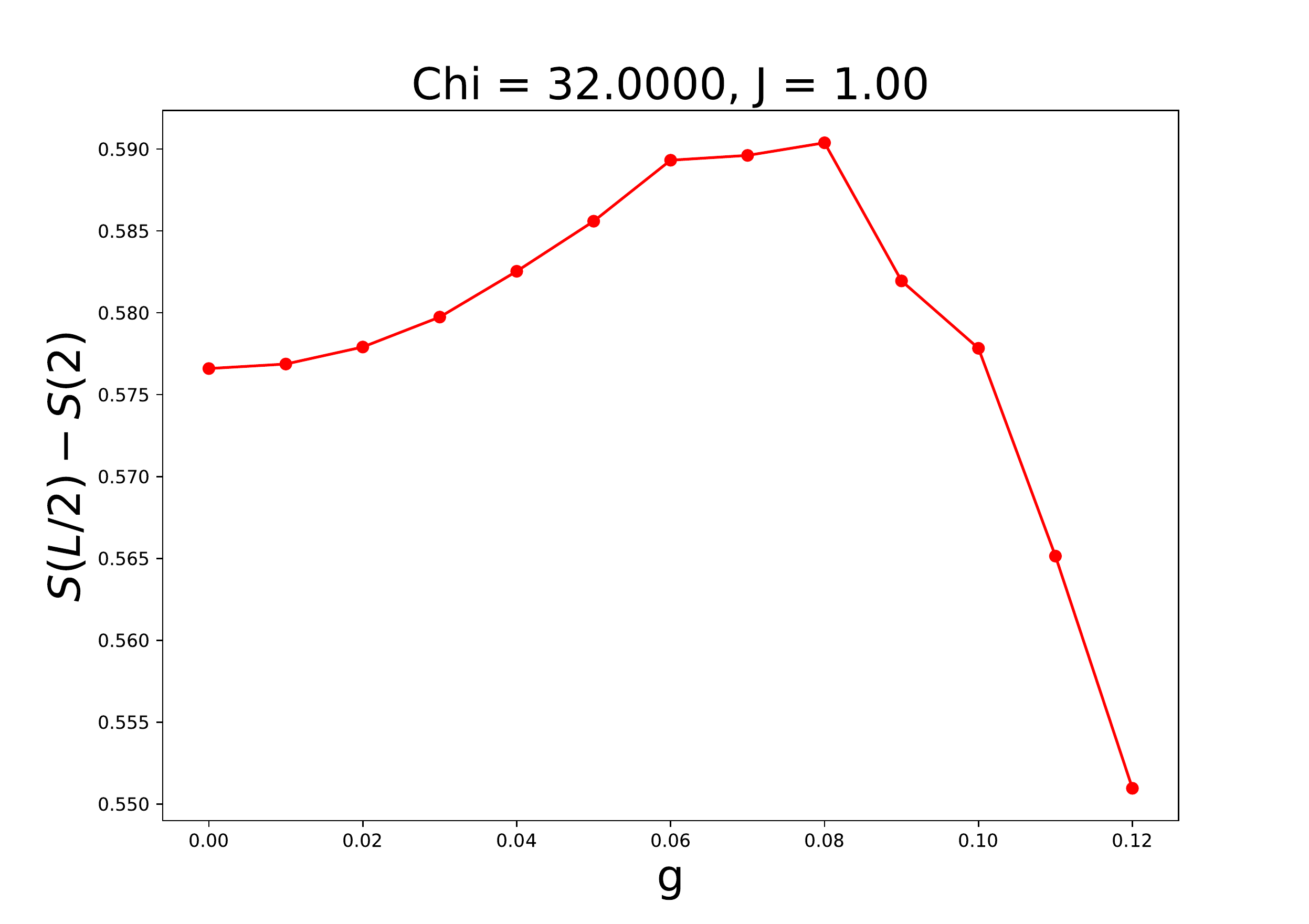}
\end{center}
\caption{
\label{fig:ceff-vs-g}
The half-chain entanglement $S(L/2) - S(2)$ at fixed $L =52$,
as a function of $g$, averaged over up to 600 samples.
}
\end{figure}

\begin{figure}[t] \begin{center}
\includegraphics[height=.4\textwidth]{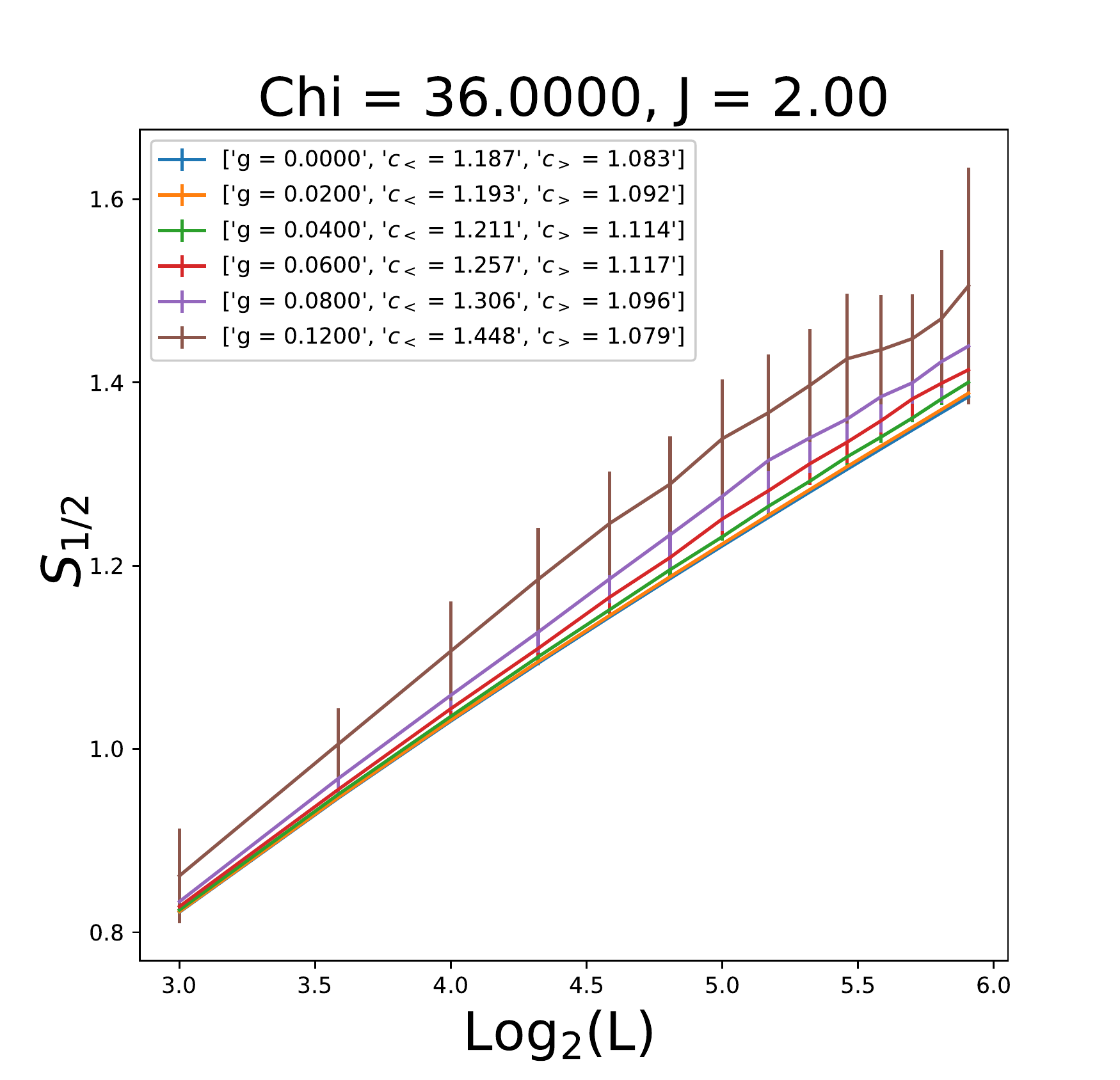}
\includegraphics[height=.4\textwidth]{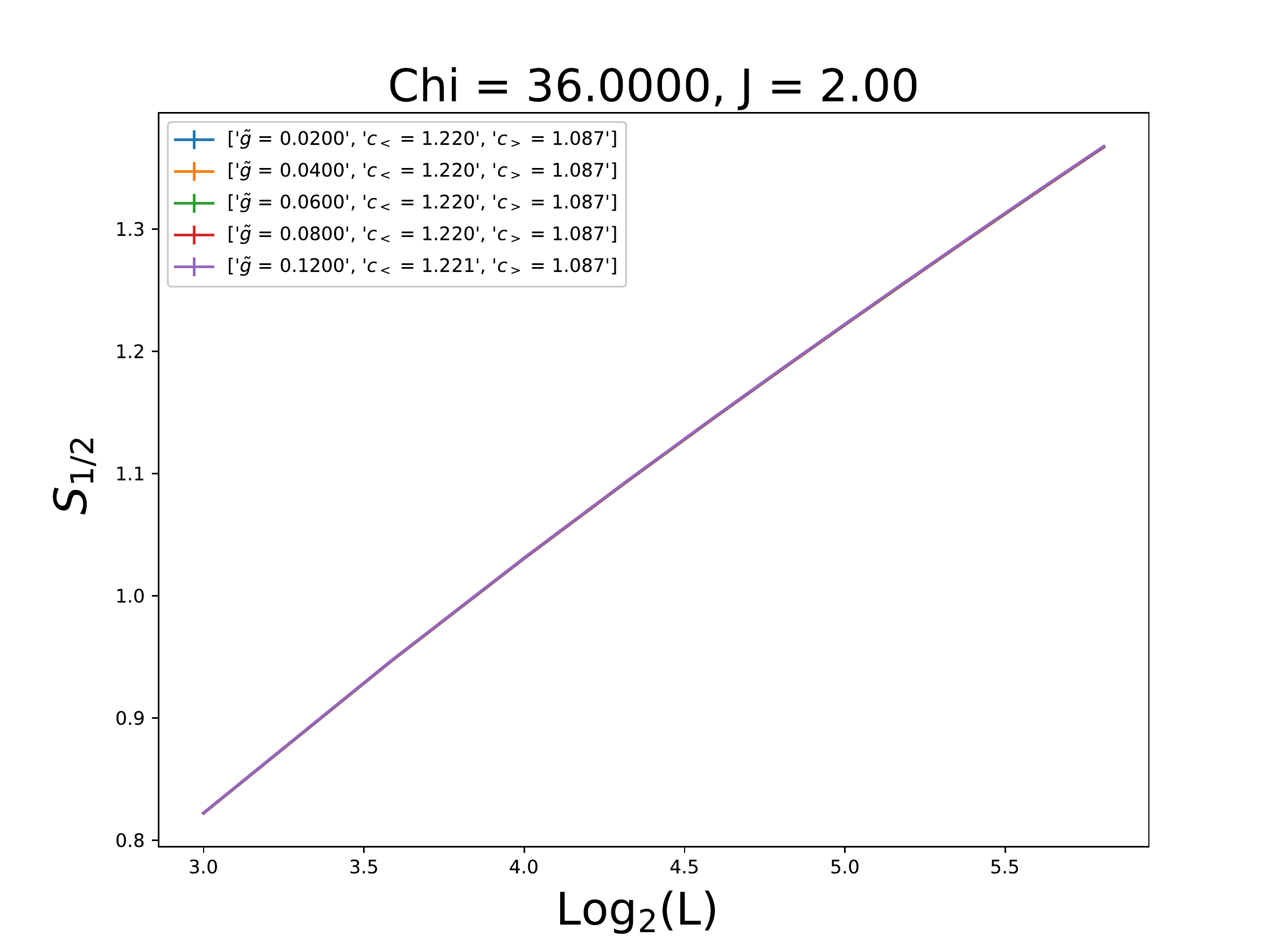}
\end{center}
\caption{
\label{fig:entropy-plots}
Left: half-chain entanglement entropy versus log of system size
for $J = 2, t = 1$, and various values of $g$, the coupling to the fermion field, computed in the truncated scheme. $c_<$ and $c_>$ are six times the slope calculated at $\log_2 L < 5$ and $\log_2 L \ge 5$ respectively.
Right: half-chain entanglement entropy versus log of system size,
for various values of $\tilde g$, the coupling to bath field.  
In the latter case, the curves all lie on top of the free fermion curve.
The inset gives fits to the slope (times 6).
}
\end{figure}

{\bf Half-chain entropies.}  
In Fig.~\ref{fig:ceff-vs-g} we plot $S(L/2)-S(2)$
at fixed $L$ for various $g$, and observe
smooth growth to a maximum value, suggestive
of scenario 3 with an intermediate-coupling fixed point. 
$S(2)$ is subtracted to remove a $g$-dependent constant shift.
Beyond the maximum, all the entanglement is destroyed;
this is the reverse Kondo phase.

The right panel of Fig.~\ref{fig:entropy-plots}
illustrates the fact that 
the coupling to bath field 
$\tilde g \psi \tilde \rchi$
is irrelevant -- 
it is identical to the free fermion answer for all $\tilde g$.

The left panel of Fig.~\ref{fig:entropy-plots}
shows the half-chain entanglement entropy 
as a function of $\log(L)$ 
for $J = 2, t = 1$, and various values of $g$, computed using the truncation scheme.
We expect these choices of $J$ and $g$ are in the regime of validity of the truncation especially for the smaller values of $g \lsim D/10$. For comparison, the results  obtained using the standard DMRG are shown in Appendix \ref{app:more_numerics}. 

There is a regime at small $g$
where the entanglement grows faster with $L$ than the free-fermion answer at small system sizes
At larger $L$, the curve levels off to approximately the same slope as the free-fermion curve. One possibility is that this is due to some extra finite-range correlation between the cluster degrees of freedom on top of the extended contribution from the itinerant degree of freedom, and that the true area law violating term has the same coefficient as a decoupled spinless fermion.

Another possiblity is that the apparent rejoining with the free fermion value is related to a previously-observed difficulty in the use of DMRG algorithms
for disordered critical systems \cite{goldsborough2017entanglement}. If we parametrize the EE as $$S_{1/2}(L) = \frac{1}{6} \log_2 L + \frac{\delta c_{dis}}{6} g(L) + const,$$ where $g(x) = \log_2 (x)$ at $x < l_{crossover} \sim 2^5$ and saturates to a constant at $x \gsim l_{crossover}$, this reproduces our observed results.  A more conservative explanation is that $c \sim 1$ throughout the extended phase.

\begin{figure}[t] \begin{center}
\includegraphics[width=.48\textwidth]{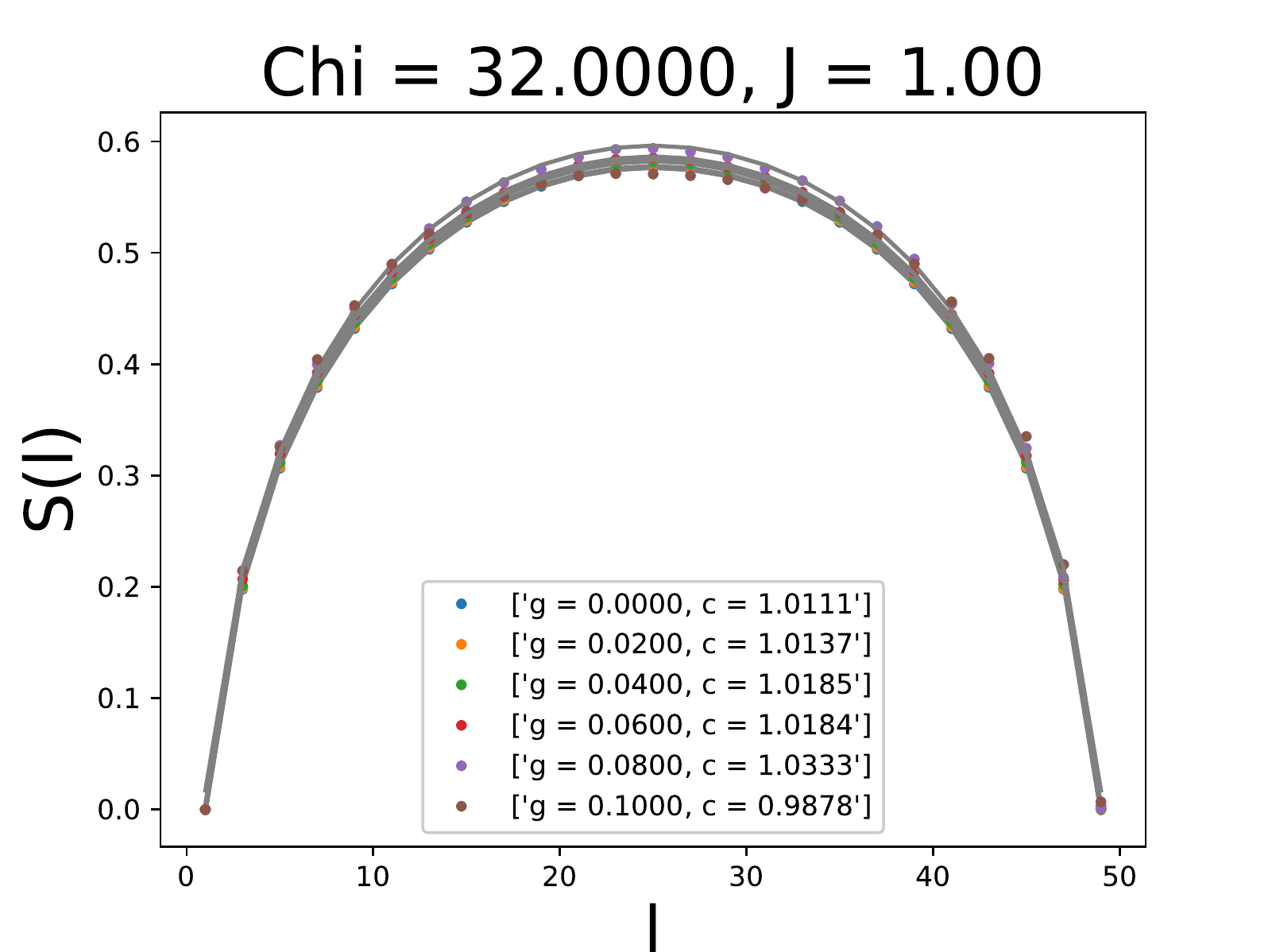}
\includegraphics[width=.48\textwidth]{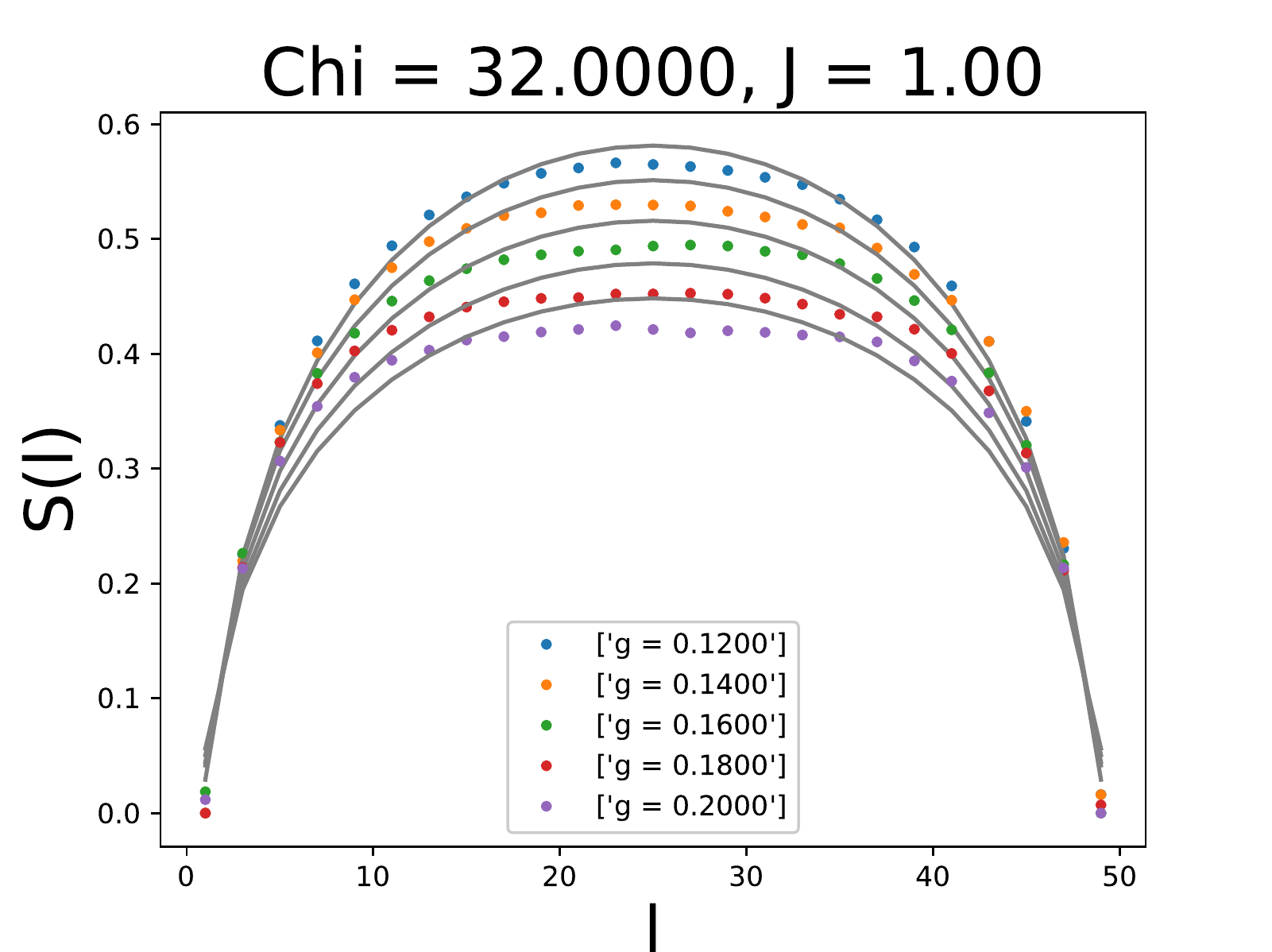}
\end{center}
\caption{
\label{fig:FSE-fit}
The finite size entanglement as a function of the position of the entanglement cut at a fixed system size $L = 52$.  We plot only the even sites in order to remove  even-odd oscillations. At small values of $g$, the data is well fit by the expression $S(x) = \frac{c}{6} \log_2 \left(\frac{2L}{\pi} \sin (\pi x/L)\right) + a$. At larger values of $g$ the fit fails. Data are averaged over $200-400$ disorder realizations. 
}
\end{figure}

In addition to the scaling of the EE with the system size, we can also look at the dependence on the size of the bipartition at fixed system size. Fitting our data to known expressions for the finite size entanglement in critical systems \cite{Holzhey:1994we,korepin2004universality,calabrese2004entanglement} provides another method to extract the `central charge'. The quality of the fit is also a useful diagnostic of whether the system is approximately critical or has a finite correlation length. Our results are shown in Figure \ref{fig:FSE-fit}; at small $g$ we observe a small rise in the entanglement but cannot draw a strong conclusion. At larger values of $g$ we observe the onset of a finite correlation length.

\begin{figure}[h] \begin{center}
\includegraphics[width=.2\textwidth]{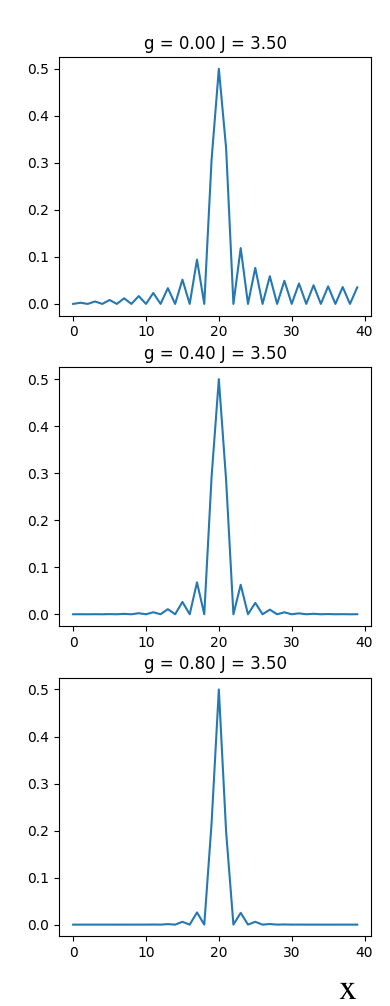}
~~~~
\includegraphics[width=.2\textwidth]{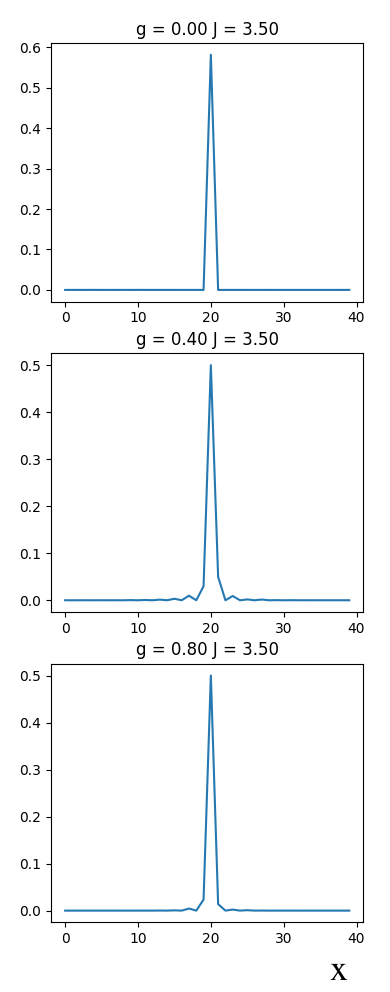}
\end{center}
\caption{
\label{fig:correlation-functions}
Left: The absolute value of correlation functions of the itinerant $\psi$ fermions
between the middle site and the site $x$, for various $g$.
Right: 
The absolute value of correlation functions of the localized $\rchi$ fermions
between the middle site and the site $x$, for various $g$.
}
\end{figure}

{\bf Correlation functions.}
Fig.~\ref{fig:correlation-functions}~shows the 
fermion equal-time correlation functions
in the DMRG approximation to the groundstate.
The absolute value is averaged over 50 instances.

The result fits well to 
\be \label{eq:fcorrelation}\bigl|\vev{ \psi^\dagger_x \psi^\nd_{L/2} }\bigr| \sim { | \sin 2 k_F (x-L/2) |\over |x-L/2|^\alpha }
\ee
with $\alpha < 1$.
The free fermion answer is of the form \eqref{eq:fcorrelation} with $\alpha=1$.
For $g>0$, the exponent is larger than the free fermion value.
It would be interesting to try to reproduce this change in the exponent
using the $q-2$ expansion.
The right column of Fig.~\ref{fig:correlation-functions} 
shows that at the same values of $g$, the so-called localized fermions $\rchi$ are indeed still localized. 
The bottom row of Fig.~\ref{fig:correlation-functions} 
shows that at large $g$, everybody is localized -- 
this is the reverse Kondo phase.

\begin{figure}[h] \begin{center}
\includegraphics[width=.6 \textwidth]{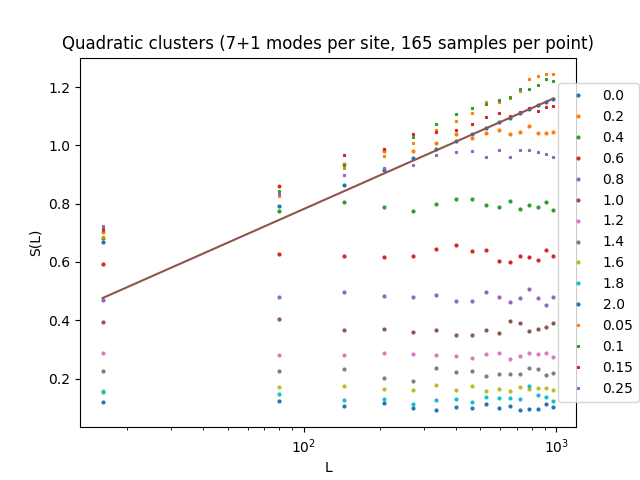}
\end{center}
\caption{
\label{fig:quadratic-clusters}
Half-chain entropy as a function of chain length
for a free-fermion chain (uniform hopping $t=1$)
hybridized with local {\it quadratic} clusters, with random quadratic intra-cluster interactions
of mean $1$.  Different curves are different values of the 
root-mean-square hybridization coupling $g$, which varies from $0$ to $2.0$.  
The solid line is the asymptotic behavior in the clean limit, $ S(L) = {1\over 6} \log(L) + a$.
}
\end{figure}

{\bf Lesion studies.}  
To what extent is the use of the SYK model 
as the cluster Hamiltonian crucial?
We can attempt to address this question
by perturbing 
the cluster Hamiltonian by (relevant) quadratic terms.
In the case of purely quadratic clusters, the entire
Hamiltonian is quadratic, and we can study larger system sizes, 
calculating the entropy by the Peschel formula \cite{peschel2003calculation}.
The result is shown in Fig.~\ref{fig:quadratic-clusters}.

Another reason to study the case of quadratic clusters
is to identify the length scale at which localization 
sets in.  In one dimension, to which our numerical 
work is sadly limited, localization is likely the inevitable 
long-distance fate. We see in Fig.~\ref{fig:quadratic-clusters} that at $g \lsim 0.3$ localization sets in at system sizes which are too large for us to accurately study using DMRG. Therefore we cannot rule out the possibility that the quartic model would show a finite correlation length as well at larger system sizes. However, whereas localization is guaranteed for the disordered quadratic system in one dimension \cite{berezinskii1995kinetics}, it is possible that the interacting system remains extended.

\section{Conclusions}

In this paper we have studied 
what happens when we couple
a Fermi surface to a lattice 
of locally critical clusters.
We have provided evidence from various 
approaches 
for the existence of a novel strange metal fixed point 
at intermediate values of the root-mean-square hybridization coupling $g$.
This fixed point is stable to perturbations of $g$.  
Intra-cluster quadratic terms are likely to
be relevant.
We note that the proposed new strange metal fixed point
is not Lorentz invariant.


A comment about the role of large $N$ is in order.  
The power-law in the SYK fermion Green's function 
is a crucial ingredient in the construction.
Such critical behavior in a (0+1)-dimensional system requires 
a large number of degrees of freedom:
if one takes $ \omega \to 0$ before
$N\to \infty$, the low-lying level spacing of the clusters will be discrete.
The effects of this phenomenon are visible in the top left of Fig.~\ref{fig:phase-diagram}:
if the level spacing of the SYK clusters is large compared to $t$, 
the hybridization coupling has no effect.
At leading order in large-$N$, 
the power-law in the cluster-fermion Green's function
is directly carried over into the itinerant fermion self-energy,
as in the holographic calculation.
In contrast, at finite $N$, only power-laws corresponding to
relevant perturbations
(in the sense of \S\ref{sec:finite-N})
affect the low-energy behavior of the itinerant fermions.  
The bath field $\tilde \rchi$, for example, is irrelevant for all $q$.  
This restricts the resulting states to have 
self-energy exponent $ 2\nu < 0$.

We conclude with a suggestion for a direction for progress towards 
corroborating the existence of
this fixed point and studying its properties.
The matrix product ansatz used in the DMRG study of 
\S\ref{sec:numerical-results} 
does not take advantage of all of the structure of the problem. 
In particular, the fact that the clusters do not couple directly
to each other represents a kind of `entanglement bottleneck'  -- 
any long-ranged entanglement along the chain 
necessarily passes through the itinerant fermion sites.
To take advantage of this, 
it would be useful to construct 
a variational tensor product state
with the structure of our interaction graph
shown in Fig.~\ref{fig:cluster-diagram}. It would also be interesting to try to apply an adaptation of the dMera of \cite{goldsborough2017entanglement} to answer the question regarding the scaling of the EE in our disordered system.


\appendix
\renewcommand{\theequation}{\Alph{section}.\arabic{equation}}

\section{$1/N$ corrections to the cluster-fermion propagator}

\label{app:turtles}

As promised in \S\ref{ref:large-N}, 
here we analyze
the $1/N$ correction to the propagator of the localized fermions.
Starting from the conformal SYK propagator as $\mathcal G_0$, and denoting convolution over the intermediate variable by $*$, we have
$$ \mathcal G(\omega,x-y) - \mathcal G_0(\omega) \delta_{xy} =   \frac{g^2}{N}\mathcal G_0(\omega)^2 G^0_{xy} \delta_{xy} + \frac{g^4}{N}\mathcal G_0(\omega)^3 \delta_{xy} G^0_{xz} * G^0_{zy} + \ldots$$
Fourier transforming and using results from \S\ref{ref:large-N}, we find
$$ \delta\mathcal G(k\omega) =  N^{-1} \left( g^2\mathcal G^2_0(\omega) \int \dbar k\; G^0(k\omega) + g^4 \mathcal G_0^3 \int \dbar k \;G^0(k\omega)^2 + \ldots\right)$$

$$ =  \frac{g^2}{N} \int \dbar k \mathcal G_0^2 \frac{1}{G_0(k)^{-1}  - g^2 \mathcal G_0} = \frac{g^2}{N} \mathcal G_0^2 \int \dbar k G(k).$$

$$ = \frac{g^2}{N} \sG_0^2(\omega) \int \dbar^d k \; \frac{1}{\ii \omega - \xi(k) + \ii g^2 (\pi/J^2)^{1/4} |\omega|^{-1/2} \sgn \omega}$$

{\bf Analysis of the integral.}
Thus the $1/N$ correction to the localized fermion propagator
is proportional to
\be\label{eq:def-of-D} D(\omega) \equiv \int \dbar^d k G(k, \omega)  =G(\omega; xx),
\ee
the local density of states of the itinerant fermions,
the quantity which determines the $dI/dV$ curve measured
by scanning-tunneling microscopy.  

Some difficulty arises from the UV-sensitivity of this integral: 
the answer is not a property of only the physics at the Fermi surface,
but depends also on short-distance details.
Here we will show that, 
given the form of the SYK propagator $\sG_0(\omega)$, 
the resulting $D(\omega)$
{\it vanishes} at small frequency, 
independent of those short-distance details.
Therefore, this $1/N$ correction
does not modify the leading low-frequency scaling behavior
of $\sG$, even at frequencies very small compared to $N$.

Let us parametrize $G$ as follows:
\be
\label{eq:parametrizeG}
G(k,\omega)= {1\over \omega - \xi(k) - \Sigma(\omega)}. \ee
(Note that we assume $k$-independent self-energy.)
To learn something about integrals of the form \eqref{eq:def-of-D},
consider free fermions 
with bandstructure $ \xi(k)$, in which case we have the Schwinger-Dyson equation
$$ \( - \ii \partial_t  + \xi(\ii \partial_x) \) G_{x,0}(t) = \delta^d(x)\delta(t) $$
and hence 
by Fourier transform
$$ G(k,\omega) = \int dt d^dx e^{ - \ii (kx - \omega t) } G_{x,0}(t),$$
\eqref{eq:parametrizeG} obtains with $\Sigma(\omega)=0$.
On the other hand, we also have
$$ G_{x,0}(t) \equiv \bra{\gs} c_x^\dagger(t) c_0^\nd(0) \ket{\gs}
= \int \dbar^d k \int \dbar^d q 
\bra{\gs} e^{ - \ii \omega_k t + \ii k x } c_k^\dagger c_q^\nd\ket{gs} 
= 
\int_{q \in FS} \dbar^d q 
e^{ - \ii \omega_q t + \ii q x } . 
$$
Therefore
\begin{align} D(\omega) {}=& \int \dbar^dk \int dt d^dx e^{ - \ii (kx - \omega t) } 
 {1\over V} \sum_{q \in FS} e^{ - \ii \omega_q t + \ii q x }
 \\ ={}&
 \int dt e^{ \ii \omega t }
 \int_{q \in FS} \dbar^d q 
  e^{ - \ii \omega_qt } 
\\  ={}&
 \int_{q \in FS} \dbar^d q 
\delta(\omega - \omega_q) 
 \\ ={}& 
 \int_{q \in FS} \dbar^d q 
 {\delta( q - q(\omega) ) \over \partial_q \omega} 
 =\theta( \mu - \omega) \rho(\omega) 
 \label{eq:expression-for-D}
\end{align}
which is the density of filled levels.
Notice that \eqref{eq:expression-for-D}
correctly reproduces 
$$ \int d\omega D(\omega) = \int \dbar^d q G_q(t=0) = \int \dbar^d q \bra{gs} c_q^\dagger c_q^\nd \ket{\gs}
= \int_{q \in FS} N_q = N  $$
the total number of fermions.

For example, consider the case when the Fermi level 
is near the edge of a 1d band, so that $ \rho(\omega)  = {1\over \sqrt{ \omega - 2t } }\theta(2t - \omega)$.
Let us reproduce this answer
using \eqref{eq:expression-for-D} starting from the Green's function
\eqref{eq:parametrizeG}.
Linearizing about the Fermi surface
$ k_\perp = k - k_F$, 
\be \label{eq:bandexpand} \xi(k) = - {
\mu }
+  v_F k_\perp   + \CO(k_\perp)^2\ee
 would give 
$$D(\omega) \buildrel{?}\over{=} \Omega_d  k_F^{d-1}  \int_{-\Lambda}^\Lambda { d k_\perp}
{1 \over f(\omega)-  v_F k_\perp  } 
= \Omega_d  k_F^{d-1} \log \( { f(\omega) - v_F \Lambda \over f(\omega) + v_F \Lambda } \) $$
which depends on $\Lambda$ at large $\Lambda$ --
this is the UV sensitivity we advertised above.
The answer will be different if we include 
the next term in the expansion \eqref{eq:bandexpand} about the Fermi surface:
$$ \xi(k) = -  
 \mu
+  v_F k_\perp +  t k_\perp^2 + ...
$$
since then the integral $ \sim \int^\Lambda {dk_\perp \over k_\perp^2 } $ would be finite as $\Lambda \to \infty$.
For definiteness, focus on the 1d band edge example:
that is, suppose $d=1$ and $\mu$ is near the bottom of the band so $v_F = 0$.
Then 
$$D(\omega) = 
\int_{-\infty}^\infty { \dbar k}
{1 \over \omega-  t k^2 } 
= {1 \over  \sqrt{ \omega t}  }
$$
(the contour can be closed in either half-plane).
The imaginary part is only nonzero for $ \omega<0$, 
and reproduces the divergence at the 1d band edge.

Armed with this intuition, we return to the case of interest, where
at low frequency the singular self-energy dominates, and $f(\omega) \sim \omega^{-1/2}$.
In that case (still in $d=1$ for now), 
$$D(\omega) = \int {\dbar k }  {1 \over C \omega^{-1/2} - \mu - \xi(k) } 
= \int {\dbar k }  {\omega^{1/2} \over C  - \( \mu - \xi(k)\) \omega^{1/2}} . $$
Assume that $\xi$ is a polynomial of degree $D$ in $k$, $\xi = a_0 k^D + a_1 k^{D-1}  + \cdots $ ; 
then letting $ u \equiv \omega^{1\over 2 D }k$, 
this integral at small $\omega$ is 
$$D(\omega) 
= \omega^{{1\over 2} - {1\over D}} \int {\dbar k }  {1 \over C  -  a_0 u^D +  a_1 u^{D-1} \omega^{{1\over 2 D }}  + \cdots } 
\buildrel{\omega \to 0}\over{\sim}\omega^{{1\over 2} - {1\over D}} $$
Alternatively, suppose the we are working in a lattice model,
so that the momentum integral $\int \dbar^d k$ is over a finite Brillouin zone;
in that case, $D(\omega) \buildrel{\omega \to 0}\over{\sim} \omega^{1/2}$.
In either case, we find $D(\omega) \buildrel{\omega \to 0}\over{\to } 0 $.

In general $d$, the same analysis gives
$$ D(\omega) = \omega^{1/2} \int {\dbar^dk \over \omega^\half k^D + ... } 
= \omega^{\half - {d \over 2 D} } K_d \int { u^{d-1} du \over u^{D} + \cdots } $$
where $K_d = {\Omega_{d-1} \over (2\pi)^d}$. 
The integral converges when $ D > d$, in which 
case the power of omega is ${\half - {d \over 2 D}  } > 0 $, 
and the integral vanishes as $\omega \to 0$.  
Alternatively, we can appeal to the lattice regulator: 
compactness of the Brillouin zone 
guarantees that $D(\omega) \buildrel{\omega \to 0} \over {\to} \omega^{\half} \int {\dbar ^d k } {1\over C}  $
is $ \omega^\half$ times a finite integral.

\section{Other numerical results}
\label{app:more_numerics}
\begin{figure}[h] 
$$\includegraphics[width=0.5\textwidth]{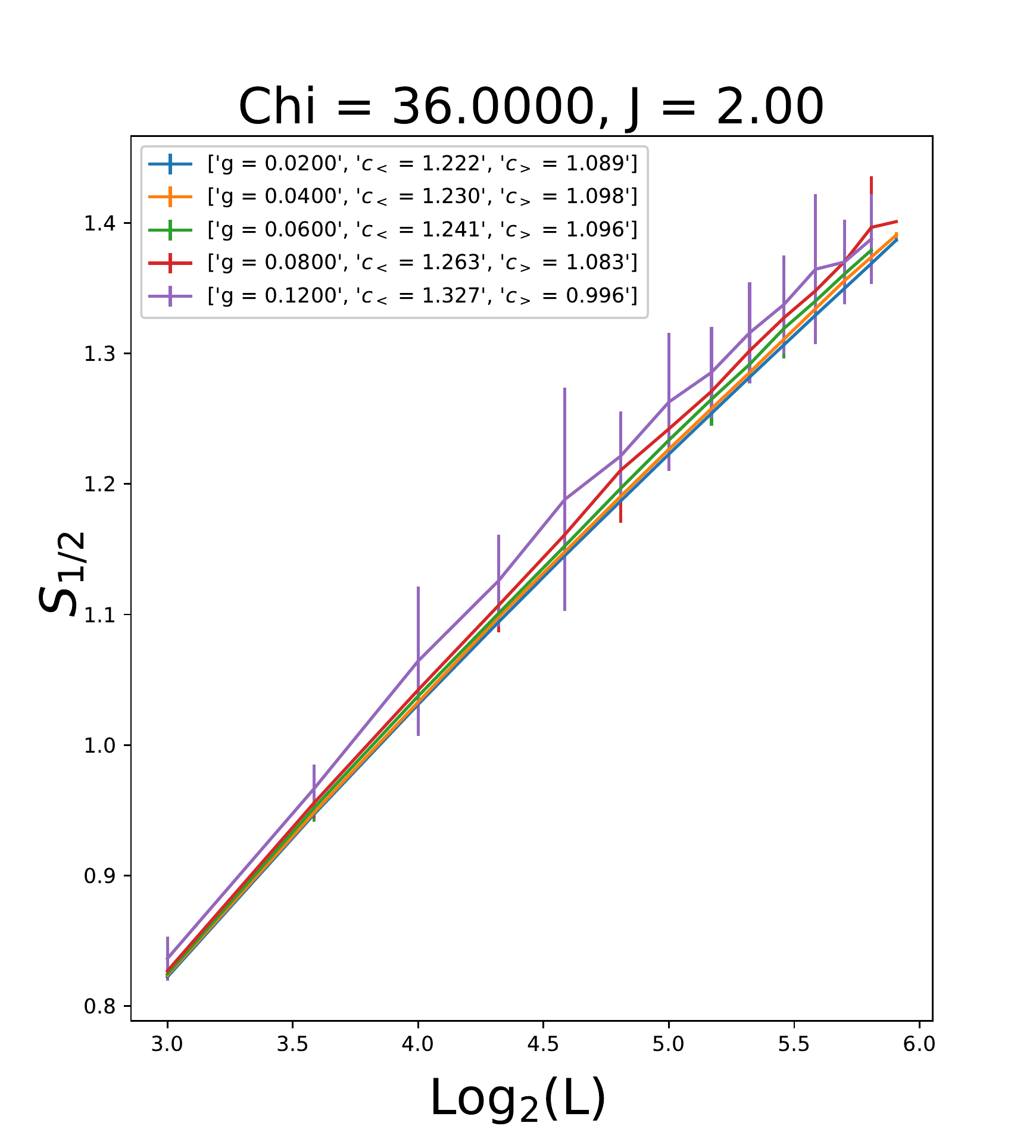}$$
\caption{$S_{1/2}$ vs Log(L) calculating using standard DMRG at small $g$.}
\end{figure}
The standard vMPS algorithm with only six cluster fermions is an especially poor representative of the large $N$ model at very small $g$; if $g$ is smaller than the finite size energy gap between the SYK ground state and the excited states then the hybridization interaction is essentially frozen out. The truncated version of the algorithm starting with a larger Hilbert space does a better job in representing the large $N$ model.

\begin{figure}[h!] \begin{center}
\includegraphics[width=.7\textwidth]{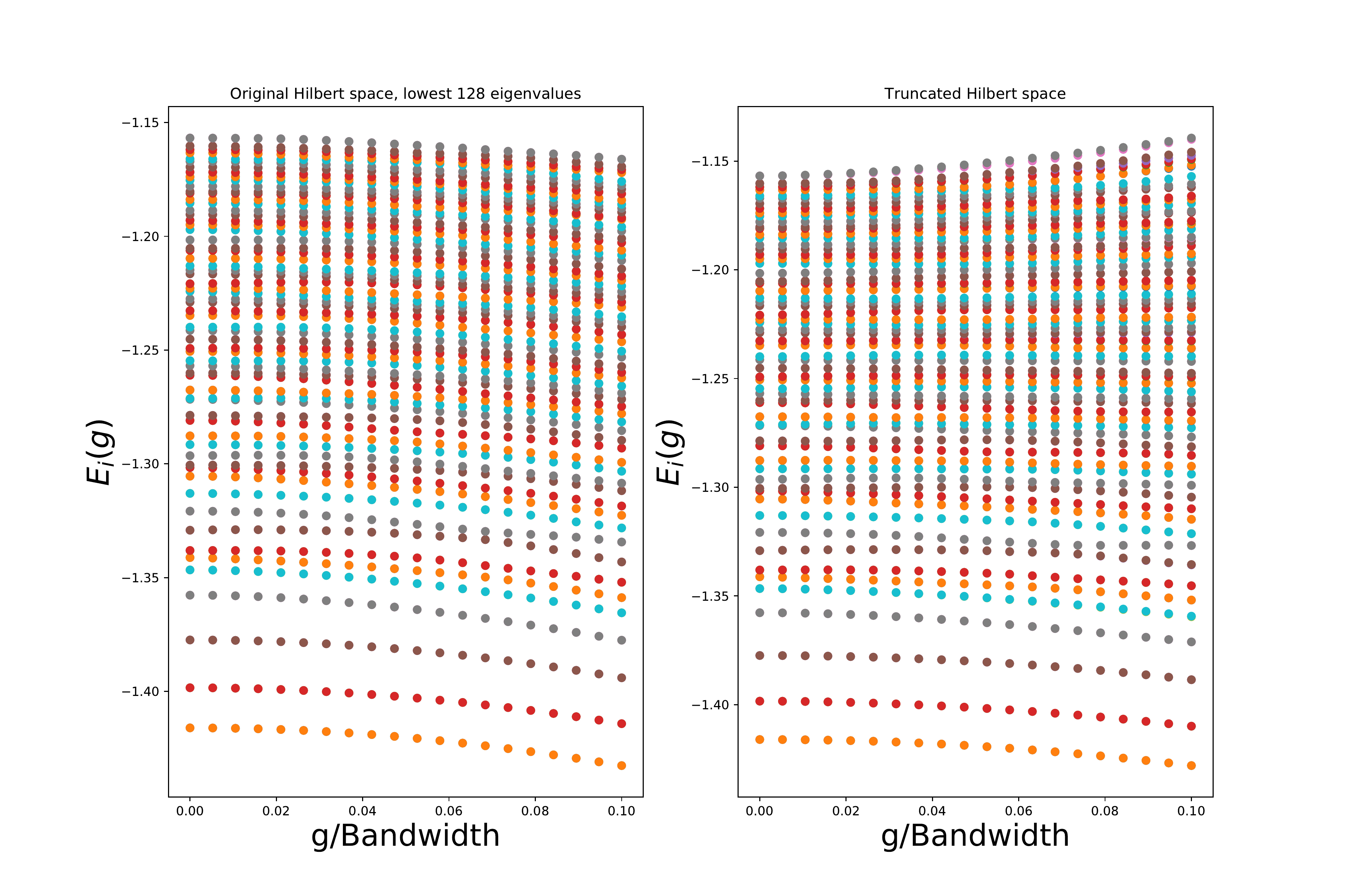}
\includegraphics[width=.7\textwidth]{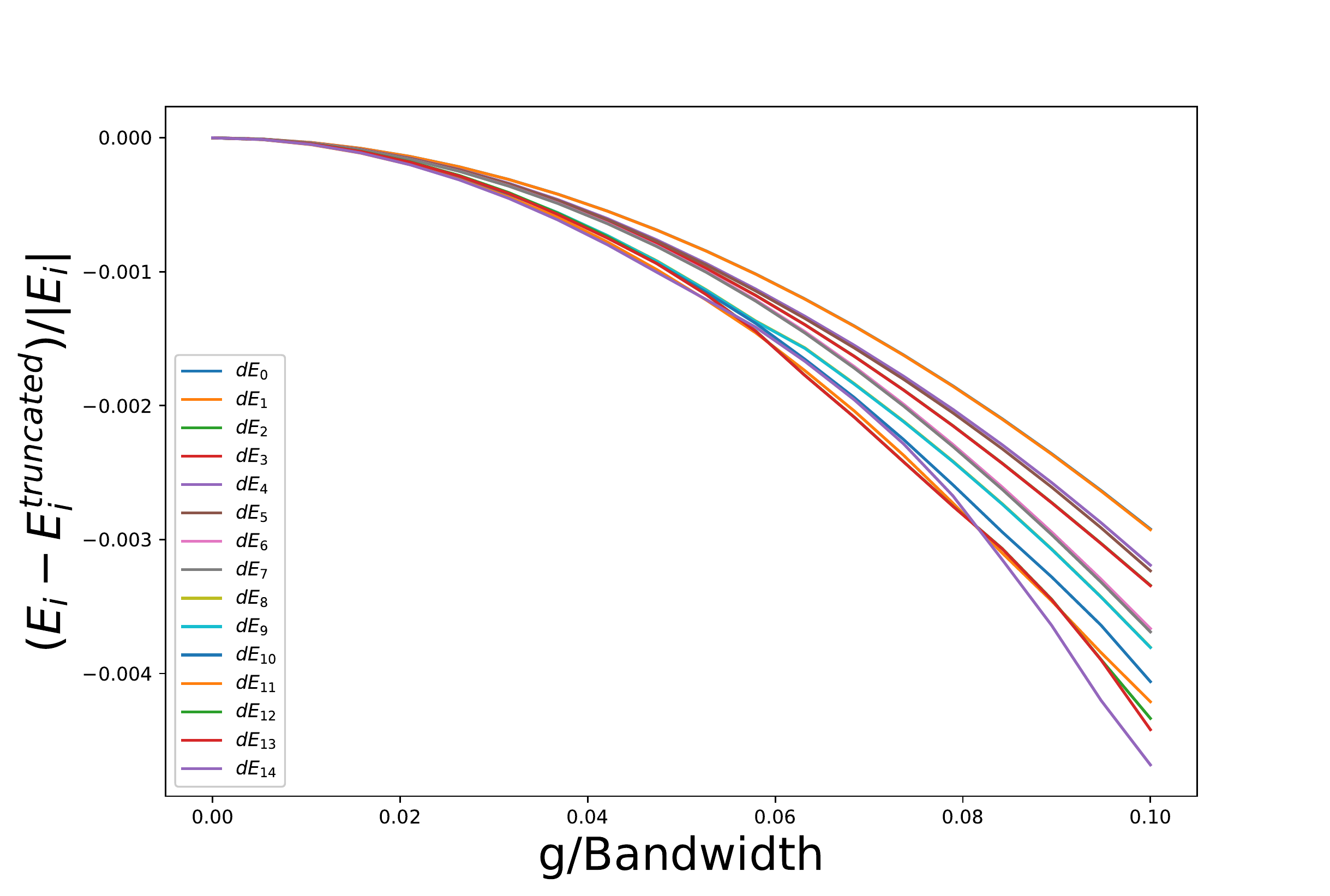}
\end{center}
\caption{
\label{fig:truncation}
A check on the validity of the truncation method.
Top Left: first 128 levels of an SYK cluster with $N_{syk}=12$ 
hybridized with a single extra fermion mode.
Top Right: the spectrum of the truncated SYK hamiltonian
(truncated to 64 levels)
coupled to an extra fermion mode. 
Bottom: Fractional error in the energy eigenvalues of the lowest fifteen states. 
}
\end{figure}

{\bf A benchmark of the truncation method.}  
The truncation method outlined above is an uncontrolled approximation 
for the sizes of local Hilbert spaces available to us.  
As a test of the method, 
in Fig.~\ref{fig:truncation}
we show the spectrum of an SYK impurity
coupled to a single extra fermionic mode (one site of the chain).
The bottom part of the truncated spectrum matches quite well with 
the correct spectrum.  The top of the truncated spectrum is wrong: 
the level repulsion from the levels above is missing.   
We used this method in studying the growth of the half-chain entanglement entropy in addition to the standard MPS algorithm.

{\bf Acknowledgements}

We thank Sid Parameswaran for collaboration at the 
initial stage of this work, 
and Dan Arovas, Tarun Grover, Aavishkar Patel and Shenglong Xu for helpful input.
This work was supported in part by
funds provided by the U.S. Department of Energy
(D.O.E.) under cooperative research agreement 
DE-SC0009919.
This research was done using resources provided by the Open Science Grid \cite{pordes2007open,sfiligoi2009pilot}, which is supported by the National Science Foundation award 1148698, and the U.S. Department of Energy's Office of Science.

\vfill\eject

\bibliographystyle{ucsd}
\bibliography{collection} 
\end{document}